\documentclass[preprint,journal]{vgtc}       % preprint (journal style)

\ifpdf%                                % if we use pdflatex
  \pdfoutput=1\relax                   % create PDFs from pdfLaTeX
  \usepackage{graphicx}                % allow us to embed graphics files
  \DeclareGraphicsExtensions{.pdf,.png,.jpg,.jpeg} % for pdflatex we expect .pdf, .png, or .jpg files
\else%                                 % else we use pure latex
  \usepackage{graphicx}                % allow us to embed graphics files
  \DeclareGraphicsExtensions{.eps}     % for pure latex we expect eps files
\fi%

\graphicspath{{figures/}{pictures/}{images/}{./}} % where to search for the images

\usepackage{microtype}                 % use micro-typography (slightly more compact, better to read)
\PassOptionsToPackage{warn}{textcomp}  % to address font issues with \textrightarrow
\usepackage{textcomp}                  % use better special symbols
\usepackage{mathptmx}                  % use matching math font
\usepackage{times}                     % we use Times as the main font
\usepackage{cite}                      % needed to automatically sort the references
\usepackage{tabu}                      % only used for the table example
\usepackage{booktabs}                  % only used for the table example
\usepackage{paralist}
\usepackage{xcolor}
\usepackage{multirow}
\usepackage{wrapfig}
\usepackage{float}
\usepackage[normalem]{ulem}
\usepackage{colortbl}
\usepackage{amsmath}
\usepackage{balance}
\usepackage{booktabs}
\usepackage{makecell}
\usepackage{dblfloatfix}

\newcommand{\etal}{\textit{et al.}}
\newcommand{\eg}{\textit{e.g.}}
\newcommand{\ie}{\textit{i.e.}}
\newcommand{\quotestudy}[1]{\textit{``#1''}}
\newcommand{\RNum}[1]{\uppercase\expandafter{\romannumeral #1\relax}}

\newcommand{\statnew}[5]{({#3})}

\newcommand{\re}[3]{\textcolor{black}{#2}}
\newcommand{\para}[1]{\vspace{1mm}\noindent\textbf{#1}}
\newcommand{\point}[1]{\emph{\underline{#1}}}

\newcommand{\numOp}{10}
\newcommand{\numInt}{18}

\newcommand{\numComb}{81} %sum of each item in cell
\newcommand{\IntOp}{interaction}

\newcommand{\numTotalIdea}{146}
\newcommand{\opFull}{data visualization command}
\newcommand{\op}{command}
\newcommand{\pIntFull}{paper action}
\newcommand{\pInt}{action}

\newcommand{\mapping}[2]{\re{\texttt{#2(#1)}}{\textit{#1$\Rightarrow$#2}}{Clarify the relationship between action-commands interactions (R1)}}

\newcommand{\dimOne}{\textit{Commands}}
\newcommand{\dimTwo}{\textit{Degree of Information}}
\newcommand{\dimThree}{\textit{Number of Paper Sheets Involved}}
\newcommand{\doi}{\textit{degree of information}}

\usepackage{xstring}
\newcommand{\numDesign}[1]{%
    \IfEqCase{#1}{%
        {}{six}%
        {num}{6}%
    }[\PackageError{numDesign}{Undefined option to tree: #1}{}]%
}

\newcommand{\task}{low-level task}

\usepackage{enumitem}
\setlist[itemize]{noitemsep, topsep=0pt}

\onlineid{1101}

\vgtccategory{Research}
\vgtcpapertype{Representations \& Interaction}

\title{Exploring Interactions with Printed Data Visualizations in Augmented Reality}

\author{Wai Tong, Zhutian Chen, Meng Xia, Leo Yu-Ho Lo, Linping Yuan, Benjamin Bach, and Huamin Qu}
\authorfooter{
\item
 Wai Tong, Leo Yu-Ho Lo, Linping Yuan, and Huamin Qu are with the Hong Kong
University of Science and Technology. E-mail: \{wtong,leoyuho.lo,lyuanaa\}@connect.ust.hk, huamin@ust.hk
\item
 Zhutian Chen is with Harvard University. E-mail: ztchen@seas.harvard.edu
\item
 Meng Xia is with Carnegie Mellon University. E-mail: mengxia@andrew.cmu.edu.
\item
 Benjamin Bach is with the University of Edinburgh. E-mail: bbach@exseed.ed.ac.uk. 
}

\shortauthortitle{Biv \MakeLowercase{\textit{et al.}}: Global Illumination for Fun and Profit}
\abstract{This paper presents a design space of interaction techniques to engage with visualizations that are printed on paper and augmented through Augmented Reality. Paper sheets are widely used to deploy visualizations and provide a rich set of tangible affordances for interactions, such as touch, folding, tilting, or stacking. At the same time, augmented reality can dynamically update visualization content to provide \textit{commands} such as pan, zoom, filter, or detail on demand. This paper is the first to provide a structured approach to mapping possible actions with the paper to interaction commands. This design space and the findings of a controlled user study have implications for future designs of augmented reality systems involving paper sheets and visualizations. Through workshops (N=20) and ideation, we identified 81 interactions that we classify in three dimensions: 1) \textit{commands} that can be supported by an interaction, 2) the specific \textit{parameters} provided by an (inter)\textit{action} with paper, and 3) the \textit{number} of paper sheets involved in an interaction. We tested user preference and viability of 11 of these interactions with a prototype implementation in a controlled study (N=12, HoloLens 2) and found that most of the interactions are intuitive and engaging to use. We summarized interactions (e.g., tilt to pan) that have strong affordance to complement ``point'' for data exploration,
physical limitations and properties of paper as a medium, cases requiring redundancy and shortcuts, and other implications for design.
}

\CCScatlist{ % not used in journal version
 \CCScat{K.6.1}{Management of Computing and Information Systems}%
{Project and People Management}{Life Cycle};
 \CCScat{K.7.m}{The Computing Profession}{Miscellaneous}{Ethics}
}

\teaser{
  \centering
  \includegraphics[width=0.8\linewidth]{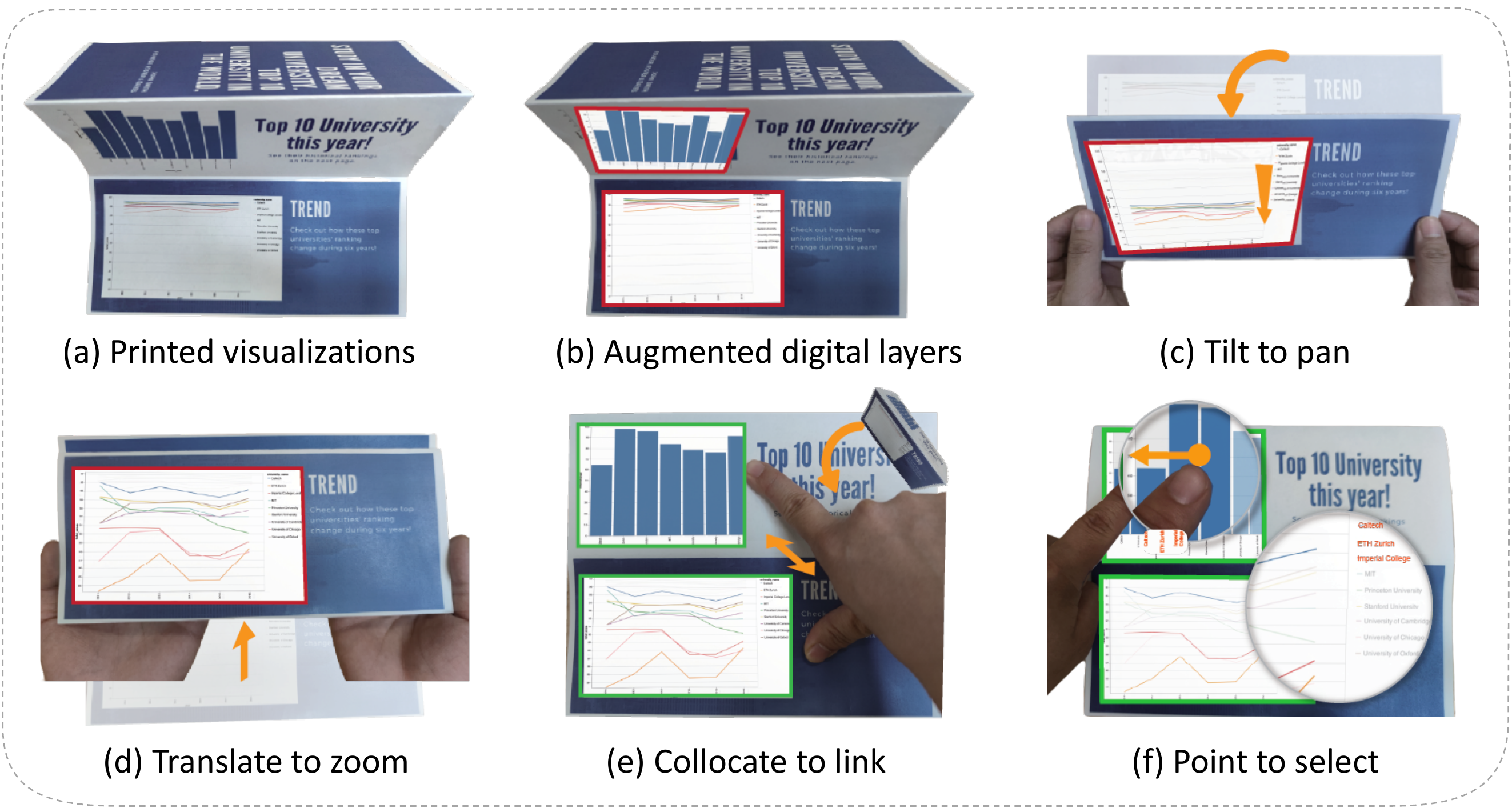}
  \caption{
   We investigate the possibility of interacting with printed visualizations in Augmented Reality. Suppose a student receives (a) a leaflet about university ranking and wants to analyze three universities' ranking history of interest.
   Examples of interactions with (b) digital content overlaid: 
(c) tilt the paper to rescale the y-axis, 
(d) move (translate) to zoom
(e) unfold to show two charts side by side and link them,
(f) point to select elements and highlight them in the other chart.
  }
  \label{fig:teaser}
}

\vgtcinsertpkg

\begin{document}

\maketitle

% \firstsection{Introduction}

%% \section{Introduction} %for journal use above \firstsection{..} instead
% This template is for papers of VGTC-sponsored conferences such as IEEE VIS, IEEE VR, and ISMAR which are published as special issues of TVCG. The template does not contain the respective dates of the conference/journal issue, these will be entered by IEEE as part of the publication production process. Therefore, \textbf{please leave the copyright statement at the bottom-left of this first page untouched}.

\section{Introduction}
% \label{sec:intro}

Interaction with visualizations is necessary for exploration, personalization, and wider engagement with data visualizations. Nevertheless, the specific means and their effectiveness for visualizations still cause considerable controversies and open research questions~\cite{hornbaek2019,dimara2019interaction}. For example, visualizations may support direct manipulation~\cite{shneiderman1981direct} through pan\&zoom, interactive lenses~\cite{Tominski2014ASO}, or brushing\&linking. However, these interactions are limited by the current interaction modalities (\eg, mouse, keyboard, touchscreen) and by the visibility and understandability of their interaction affordances~\cite{boy2016suggested}.
Unnatural interaction, unnoticed affordances, repetitive interactions, ambiguous interaction goals~\cite{amar2005low}, or missing general interaction literacy~\cite{Bach2019ceci} pose serious obstacles to people engaging with data through visualizations.

To improve affordances and provide for effective interaction, different interaction modalities have been explored~\cite{lee2018multimodal}.
For example, natural language interaction uses voice as a medium for interaction to support querying and creation of visualizations~\cite{cox2001multi}; data physicalizations provide affordances through three-dimensionality, situatedness, tangibility~\cite{jansen2013interaction}, and even dynamicity~\cite{taher2015exploring}. Recently, virtual and augmented reality further provide the potential for display and direct interaction~\cite{cordeil2017imaxes,ens2020grand} as well as offer combinations and hybridizations with tangible means for visualization and interaction~\cite{cordeil2017design,cordeil2020embodied,bach2017hologram}.

\re{Complementing this line of research, we explore \textit{paper sheets} as means to interact with visualizations printed onto these paper sheets.}{Complementing this line of research, we explore \textit{paper sheets} as tangible means to interact with visualizations printed onto these paper sheets under augmented reality.}{"Complementing this line of research..." - I see several lines of research discussed here.} We are interested in how far papers can provide affordances and means for direct manipulation with visualizations, and how to inform building systems that use these interactions.
This research is motivated by paper being a cheap means to distribute and access information through, \eg, infographics~\cite{chen2020augmenting}, newspapers, posters, books, data comics~\cite{zhao2015data,bach2017emerging}, or zines~\cite{McNutt2021zines}.
% Paper supports high-resolution and potentially large-sized formats and provides easy access.
Augmented reality (AR)---supported through camera-bearing mobile devices or Head-mounted displays (HMD)---can update such static visualizations~\cite{chen2020augmenting} and bring interactivity to them by overlaying digital layers. While previous work has demonstrated interaction techniques to interact with data visualization in AR~\cite{cordeil2017design,buschel2018interaction}, \textit{paper} upon which the visualization is printed provides its very own affordances for interaction. These techniques are only marginally explored yet~\cite{spindler2010tangible,kim2012embodied,bach2017hologram}. For example, paper can serve as a touch surface, which can be moved and rotated in space, bent, folded, moved, tilted, or stacked onto other paper sheets and even torn apart and crumpled. We argue that these interactions might occur naturally and require \re{little}{less}{There is a reasonable motivation in the introduction, but the paper should better motivate why you would use paper for augmented reality, where no physical tangible is needed. Why not just gestures?} training and practice to perform \re{}{than customized physical tangible devices, such as~\cite{cordeil2017design,cordeil2020embodied,smiley2021made,issartel2014slicing, ssin2019geogate, ens2020uplift}, due to familiarity}{same as previous}, yet provide an effective means to interact with the data. \re{}{Moreover, paper sheets provide tangible surfaces which ease arm fatigue compared to mid-air gestures~\cite{cordeil2017design}.}{same as previous} \re{}{With paper interactions, people could easily interact with printed visualization distributed in exhibitions and presentations. Besides, interacting with printed visualizations could be helpful in visualization education~\cite{bae2021touching} and brainstorming~\cite{subramonyam2019affinity, subramonyam2022composites}. Moreover, paper interactions could possibly facilitate casual collaborative visual analytics~\cite{ens2020uplift} due to its low technical barrier and enhance existing authoring tools~\cite{sicat2018dxr, cordeil2019iatk, chen2020augmenting, DBLP:journals/tvcg/ChenS0WQW20} to support interactive visualizations in AR using paper interactions.}{Clarify the target users and usage scenarios (R3, R4)}

% \begin{figure}
%   \includegraphics[width=\columnwidth]{figures/teaser1.pdf}
%   \centering
%   \caption{
% Examples of interactions with (a) a visualization printed on paper and (b) digital content overlaid: 
% (c) tilt the paper to rescale the y-axis, 
% (d) move (translate) to zoom
% (e) unfold to show two charts side by side and link them,
% (f) point to select elements and highlight them in the other chart.}
% \end{figure}
In this work, we present a design space of possible interactions with paper sheets and visualizations enhanced through AR. This design space helps us analyze interactions, inform future interfaces, and point to open research questions. For the purpose of this paper, we define \textbf{an \textit{interaction} as the mapping of an \textit{action} onto a \textit{command}, which we denote as a function \mapping{action}{command}.}
% An instrument decomposes interaction into two layers: the 
% interaction between the user and the instrument, defined as 
% the physical action of the user on the instrument and the 
% reaction of the instrument and the interaction between the 
% instrument and the domain object, defined as the command 
% sent to the object and the response of the object, which the 
% instrument may transform into feedback to the user. The 
% instrument is composed of a physical part, the input device, 
% and a logical part, the representation of the instrument in 
% software and on the screen.
Using terminology from the instrumental interaction framework~\cite{Beaudouin-Lafon}, an \textit{action} is any manipulation applied to an \re{interface}{instrument}{Clarify the terminology from instrumental interaction and apply it consistently (R1)} (\eg, a \textit{point}, \textit{rub}, or \textit{fold} to a paper sheet) while a \textit{command} is \re{the result of any action}{an interaction task applied to a domain object}{Clarify the terminology from instrumental interaction and apply it consistently (R1)} (\eg, \textit{pan}, \textit{zoom}, and \textit{filter} \re{}{a data visualization}{Clarify the terminology from instrumental interaction and apply it consistently (R1)}). For example, we can map the action \textit{fold} to the command \textit{filter}. Consequently, we denote this interaction as \mapping{fold}{filter} (speak: \textit{``fold-to-filter''} or \textit{``filter-by-fold''}) The parameters that the \textit{fold} action provides, \eg, the degree of folding, can be used to parameterize the \textit{filter} command, \eg, define a threshold for filtering a set of elements from the visualization. Figure~\ref{fig:teaser} shows an example of how a student performs a set of interactions onto a leaflet provided on an university open day.
% \re{Figure~\ref{fig:teaser} shows an example of how a user performs a set of interactions onto a paper.}{Figure~\ref{fig:teaser} shows an example of how a student performs a set of interactions onto a leaflet provided on an university open day. Besides public events, interacting with printed visualizations could be helpful in visualization education~\cite{bae2021touching} and brainstorming~\cite{subramonyam2019affinity, subramonyam2022composites}. Moreover, paper interactions could possibly facilitate casual collaborative visual analytics~\cite{ens2020uplift} due to its low technical barrier and enhance existing authoring tools~\cite{sicat2018dxr, cordeil2019iatk, chen2020augmenting, DBLP:journals/tvcg/ChenS0WQW20} to support interactive visualizations in AR using paper interactions.}{Clarify the target users and usage scenarios (R3, R4)}

We collected \numTotalIdea{} ideas, \numOp{} commands, and \numInt{} actions from both an extensive literature survey and an ideation workshop with 20 participants (graduate students and researchers in visualization and HCI) (Section~\ref{sec:workshop}). Then, we
extracted \numComb{} \IntOp{}s from these ideas and constructed a three-dimensional design space to classify interactions and guide the design of future \IntOp{}s. The dimensions include
1) the commands supported by an interaction (\eg, \textit{zoom}, \textit{pan}, \textit{filter}, etc.),
2) the specific parameters provided by an interaction (\textit{boolean}, \textit{position/area}, \textit{direction+value}, and \textit{free expression}), and 
3) the number of paper sheets involved in an interaction (\textit{1} or \textit{many}). Each interaction, being a combination of an action and a command, can be classified along these three dimensions.

Selecting 11 interactions by focusing on those \op{}s used for view manipulation as described by Heer and Shneiderman~\cite{heer2012interactive}, we then built an experimental prototype using HoloLens 2 and ran a user study (Section~\ref{sec:eval}) with 12 participants. We were interested in participants' subjective considerations (preference, comfort, intuitiveness, and engagement) as well as interactions' practical viability by observing possible combinations and confounds when using multiple interactions in the same system. Our selected interactions involve eight \pInt{}s and four \op{}s (\ie, select an interval, zoom, pan, and link\&select).

Participants were highly positive towards paper interactions and engaged with the techniques, seamlessly using \pIntFull{}s to explore static visualizations. We summarize our main findings into \numDesign{} design implications that can inform future designs for interacting with visualizations on paper in AR (Section~\ref{sec:discussion}). For example, designers can consider alternative \IntOp{}s (\eg, \mapping{tilt}{pan})
when one \IntOp{} (\eg, \mapping{point}{pan}) faces technical barriers or is not optimal for different data exploration purposes (\eg, casual exploration).
All materials from the user studies and ideation workshops, and a demo video of the experiment prototype can be found at \url{https://paperinteraction.github.io}.

\section{Related Work}
% \label{sec:related}
%This section reviews interaction techniques for data visualization in AR,\pIntFull{} techniques \& their implication for the Visualization community (VIS), and visualization task taxonomies.

% place it here to avoid skipping page: https://tex.stackexchange.com/questions/167186/figure-environment-skips-page-while-using-two-column-document
\begin{figure*}[t]
 \centering % avoid the use of \begin{center}...\end{center} and use \centering instead (more compact)
 \includegraphics[width=\textwidth]{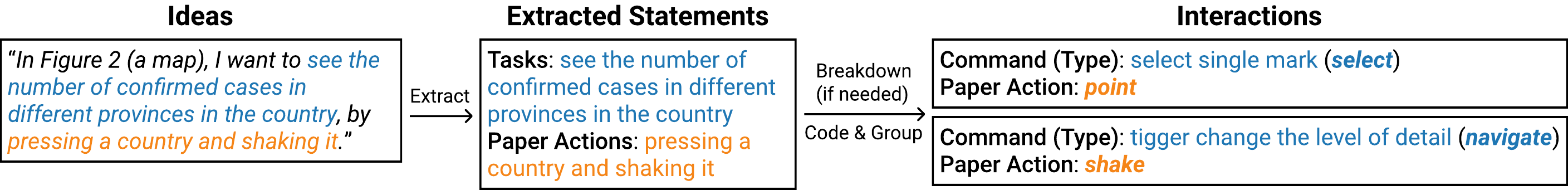}
 \caption{This figure illustrates the analysis procedure of the ideas collected from the workshop.}
 \label{fig:analysis_procedure}
%  \vspace{-3mm}
\end{figure*}

\para{Interaction Techniques for AR Visualization.}
Existing interaction techniques for data visualizations under AR can be roughly classified into five categories based on their modalities: \textit{mid-air}, \textit{tangible}, \textit{touch}, \textit{gaze and speech}, and \textit{spatial} based interfaces.
Given that mid-air hand gestures are natural and intuitive for general users,
several works have adopted mid-air hand gestures to help users navigate maps~\cite{satriadi2019augmented} and static visualizations projected on projector screens~\cite{kim2017visar}.
However, mid-air hand gestures can cause arm fatigue~\cite{hincapie2014consumed} and thus are not suitable for long-term usage.
To ease the arm fatigue issue~\cite{cordeil2017design},
researchers have built AR tangible interfaces~\cite{billinghurst2008tangible}.
For example, tangible objects, such as paper cards~\cite{spindler2010tangible, bach2017hologram,kim2012embodied}, 
paper spheres~\cite{englmeier2020tangible},
embodied axes~\cite{cordeil2020embodied,smiley2021made}, and
custom widgets~\cite{issartel2014slicing, ssin2019geogate, ens2020uplift} have been utilized as controllers for users to interact with AR visualizations.
As an alternative to the tangible interface, the touch interface on physical objects could be used to manipulate digital information precisely~\cite{cordeil2017design,elsayed2016blended,benko2007balloon,butscher2018clusters}. 
For example, the touch interface on the tabletop has been utilized for 3D parallel coordinate plot specification and manipulation~\cite{butscher2018clusters} and 3D selection~\cite{benko2007balloon} in AR.
Xiao~\etal~\cite{xiao2018mrtouch} further proposed turning every flat surface into a touch screen for head-mounted mixed reality systems.
Furthermore, researchers started to utilize gaze and speech interfaces for data visualization interaction~\cite{kim2017visar,hubenschmid2021stream,mahmood2019improving} because AR HMDs natively support these interactions (\eg, Microsoft HoloLens 2 and Magic Leap 1).
Lastly, spatial user interfaces are also increasingly used for data navigation and placement. For example, researchers extended mobile devices as spatial devices to facilitate 3D data navigation~\cite{buschel2019investigating} and visualization placement in the 3D environment~\cite{hubenschmid2021stream}.

Nevertheless, \textit{paper}, leveraging the benefit of touch ability, unique tangibility, and spatial interface, is only marginally explored~\cite{spindler2010tangible, bach2017hologram,kim2012embodied} for data interaction, especially when visualizations can be easily printed on paper.
As such, we aim to explore how to utilize the ``already there'' paper to manipulate data visualizations printed on paper directly.

\para{Paper Interactions in HCI and Visualization.}
\label{ssec:related:paper}
Previous work in the Human-Computer Interaction (HCI) literature investigated the use of paper and its metaphors to achieve better interaction design for tangible interfaces~\cite{gupta2020replicate, holman2005paper}, desktops~\cite{agarawala2006keepin}, and touchscreens~\cite{girouard2012displaystacks, strohmeier2016reflex, lahey2011paperphone, terrenghi2007affordances}. 
For example, Holman~\etal~\cite{holman2005paper} proposed eight paper gestures for interacting with the digital information projected on paper. They are \textit{hold}, \textit{collocate}, \textit{collate}, \textit{flip}, \textit{rub}, \textit{staple}, \textit{point}, and \textit{two-handed pointing}.
% benefits of paper
Utilizing paper interaction has been found to make interaction design more playful and enjoyable, and further help users leverage real-world knowledge in performing the proposed interactions~\cite{agarawala2006keepin}.

% its Implication in VIS
At the same time, researchers in the visualization community have also used natural interactions with the paper to create more effective ways to interact with visualizations.
For example, papers can be utilized as an extra layer on top of a tabletop to interactively display more information~\cite{spindler2010tangible,kim2012embodied},
while Bach~\etal~\cite{bach2017hologram} used paper cardboard to interact with three-dimensional holograms.
Spindler~\etal~\cite{spindler2010tangible} further summarized a set of interaction vocabularies for tangible views, such as \textit{translation} and \textit{rotation}.
Besides using the paper as a planar, paper could also be used as a prop to interact with 3D visualization of thin fiber structures~\cite{jackson2013lightweight} and a printed wheel chart to interact with volume visualization~\cite{stoppel2016vol}.
Moreover, these paper interactions and their metaphors (\eg, \textit{piling} and \textit{folding}) are heavily used in the traditional desktop and mobile environment for data visualization tasks, such as
comparison~\cite{tominski2012interaction},
navigation~\cite{elmqvist2008melange},
organization~\cite{lekschas2020generic, bach2015small}, 
coordination~\cite{langner2017vistiles},
and set operations~\cite{sadana2014onset}.
Different from utilizing paper interactions as metaphors in the desktop environment,
we explore the possibility of using paper as a touch, tangible, and spatial interface for people to interact with digital data intuitively and engagingly in the physical world through AR. We construct a design space to provide designers with a structured way to design systems using paper interactions as well as designing further paper interactions.

\para{Visualization Task Taxonomies.}
\label{ssec:related:task}
Many works have summarized data exploration tasks as high-level tasks~\cite{liu2010mental,brehmer2013multi,lam2017bridging} and \task{}s~\cite{yi2007toward, heer2012interactive, brehmer2013multi}.
High-level tasks, such as \textit{identify}, \textit{compare}, or \textit{summarize}, describe \textit{why} users interact with a visualization~\cite{brehmer2013multi}. 
% Low-level tasks.
Since low-level tasks are building blocks for high-level tasks~\cite{brehmer2013multi,heer2012interactive,yi2007toward}, we focus on how paper interactions support low-level tasks.
Yi~\etal~\cite{yi2007toward} proposed a set of seven low-level tasks, for instance, \textit{select}, \textit{filter}, and \textit{connect}.
Later, Heer and Shneiderman~\cite{heer2012interactive} further suggested twelve low-level tasks for data \& view specification (\ie, \textit{visualize}, \textit{filter}, \textit{sort}, and \textit{derive}), view manipulation (\ie, \textit{select}, \textit{navigate}, \textit{coordinate}, and \textit{organize}), 
and analysis process (\textit{record}, \textit{annotate}, \textit{share}, and \textit{guide}). % \& provenance
Besides, Brehmer and Munzner~\cite{brehmer2013multi} added \textit{change} to the low-level tasks.

However, these tasks are mainly explored and summarized in the desktop environment and thus actions beyond the use of the mouse and keyboard are seldom discussed~\cite{dimara2019interaction,jansen2013interaction,lee2012beyond}.
Our work utilizes the existing low-level tasks as an initial set of \textit{\op{}s} to explore how \textit{\pInt{}s} on paper sheets can be used to execute these \textit{\op{}s}.
\section{Soliciting Interactions}
\label{sec:workshop}

% \begin{figure*}[t]
%  \centering % avoid the use of \begin{center}...\end{center} and use \centering instead (more compact)
%  \includegraphics[width=\textwidth]{figures/analysis procedure.pdf}
%  \caption{This figure illustrates the analysis procedure of the ideas collected from the workshop.}
%  \label{fig:analysis_procedure}
% %  \vspace{-2mm}
% \end{figure*}

To understand the potential of using paper as an interaction medium for data exploration, we conducted an ideation study.
% With the instrumental interaction framework~\cite{Beaudouin-Lafon},
% the process of interacting with an augmented printed visualization can be described as follows:
% a user performs an \emph{\pInt{}} with the paper sheet to execute a \emph{\op{}} to the visualization.
Based on the existing literature survey as mentioned in
% Section~\ref{ssec:related:paper} and Section~\ref{ssec:related:task},
Section 2,
we start exploring possible \IntOp{}s with \pIntFull{}s: \textit{hold}, \textit{collocate}, \textit{collate}, \textit{flip}, \textit{rub}, \textit{staple}, \textit{point}, and \textit{two-handed pointing}~\cite{holman2005paper},
and \opFull{}s: \textit{visualize}, \textit{filter}, \textit{sort}, \textit{derive}, \textit{select}, \textit{navigate}, \textit{coordinate}, \textit{organize}, and \textit{change}~\cite{brehmer2013multi,heer2012interactive}.

\subsection{Ideation Workshop}
%To gather more potential mappings between \pIntFull{} and \opFull{} from the users' perspective, we conducted an ideation workshop.

\para{Participants:}
We invited 20 researchers (one Associate Professor, two Postdoctoral fellows, and 17 Ph.D. students, aged between 22 and 30; 15 males and 5 females) \re{}{14 participants came from the same research lab as the first author, four participants joined from other research labs in the same university, and two from other universities}{It would be nice to know the relationship between the workshop participants and the researchers. Are they part of the same department or group? This does not have implications toward the result, but can increase the transparency of the work.}. We selected participants with VIS or HCI backgrounds to provide more detailed ideas and start brainstorming in a shorter time due to familiarity with visualization and interaction design~\cite{bressa2019sketching}.

\para{Setup and Materials:}
To encourage participants to brainstorm more creative ideas (other than familiar point-related gestures), we divided participants into groups of four and across five sessions, inspired by the \textit{partners} technique~\cite{morris2014reducing}.
We constructed five basic charts (bar chart, pie chart, line chart, scatter chart, and choropleth map) with a Covid-19 dataset (ending on January 17\textsuperscript{th} of 2021) from data repository\footnote{\url{https://github.com/CSSEGISandData/COVID-19/}} (containing two-dimensional data, temporal data, and spatial data) to cover the common visualizations and data types encountered in daily life.
Since \pIntFull{} techniques can involve multiple papers/visualizations with the same chart type (\eg, stacking one bar chart on another bar chart), we provided participants two sets of five charts with the confirmed and recovery datasets of Covid-19 cases.
Moreover, since the size of the visualization may affect the \pIntFull{}, we printed two versions of each chart: A4 width and half-A4 width.
In total, each participant received 20 ($5\times2\times2$) charts.
Due to the Covid-19 pandemic, all sessions were hosted on Zoom. Participants were asked to print out these materials before the sessions. Each participant received \$10 for compensation.

\para{Procedure:}
Each session lasted about 90 minutes, consisting of three parts: introduction (15 mins), individual brainstorming (20 mins), and group discussion (55 mins).
In the introduction, we first briefly introduced the project background and the workshop's goal.
To encourage participants to produce diversified ideas on \IntOp{}s,
we provided two demonstrations created by the authors (flipping a sheet of paper to trigger filtering, and collating two papers to combine two bar charts into one grouped bar chart), inspired by the \textit{priming} technique~\cite{morris2014reducing}.
% The introduction was followed by a Q\&A session that allowed participants to ask any questions they might have.
Then, we gave participants a task and asked them to spend 20 minutes brainstorming the \op{}s they would like to perform on the printed visualizations and how they would achieve these \op{}s by interacting with the paper. The task description was \textit{Given static Covid-19 figures from a report, what do you want to know more about from the visualizations printed on paper and how will you interact with them?}.
To \re{help participants brainstorm novel ideas}{accelerate and encourage participants to brainstorm novel ideas}{However, the initial set of commands and paper actions were provided to the users. Somehow users feedback is limited in case all conditions have been provided. They are only allowed to make comments based on the pre settings. Why did not let them to generate actions and arrange the potential actions to different categories (Boolean, position, etc).}, we provided the initial set of \op{}s and \pIntFull{}s (we excluded point-related gestures like pointing and two-handed pointing for more diverse ideas) as prompts\re{}{, adapted from~\cite{bressa2019sketching}. We also encouraged participants to generate ideas beyond the actions mentioned in the list}{Same as above}.
Participants were then asked to write down their thoughts without considering any technological restrictions and send them to the host. The host then organized all ideas in a Google Doc for later group discussion.
After the individual brainstorming step, each participant shared and demonstrated their ideas to the group.
The group then discussed the ideas and brainstormed more ideas (\ie, build upon each other's ideas) and usage scenarios based on these individual ideas in the Google Doc for 55 minutes. As we were interested in collecting a wide variety of ideas for our design space, we did not seek a consensus for a single ``ideal'' mapping between \pInt{} and \op{} at the end of each session.
All sessions were recorded.

\subsection{Data Analysis Procedure}
\label{ssec:analysis}

In total, we have gathered \numTotalIdea{} ideas from both individual brainstorming and group discussions in all sessions.
To extract \IntOp{}s from these ideas, we performed the following analysis procedures (illustrated in \autoref{fig:analysis_procedure}).
First, the lead author extracted statements involving tasks and \pIntFull{}s from the ideas and broke down statements into multiple single \pInt{} and task mappings if necessary.
% to better assign the mapping between \op{} and \pInt{}.
Next, two authors independently coded the type of \op{}s and \pInt{}s for the mappings in the first two sessions according to the initial set of \opFull{}s~\cite{heer2012interactive,brehmer2013multi} and \pIntFull{}s~\cite{holman2005paper, spindler2010tangible}.
For the \op{}s and \pIntFull{}s that did not fit into the existing taxonomy,
the same two authors independently open-coded them and discussed their definitions.
For example, participants offered the ideas of \textit{folding} and \textit{tearing} the paper, which were not in the initial set of \pIntFull{}s.
The same two authors iteratively discussed and refined the coding scheme until reaching a Cohen's $\kappa$~\cite{cohen1960coefficient} above 0.7 for all classes of \pInt{}s and \op{}s.
The lead author then coded the rest of the sessions.
For each \op{} category, the lead author further
% used affinity diagramming~\cite{hartson2012ux} to
grouped \op{}s with similar meanings (\eg, ``\textit{select a country (from a map)}'' and ``\textit{select a timestamp (from a line chart)}'' are grouped to ``\textit{select single mark}'').
% \end{enumerate}
Finally, we had summarized \numComb{} unique \IntOp{}s.

\begin{figure*}[t]
 \centering % avoid the use of \begin{center}...\end{center} and use \centering instead (more compact)
 \includegraphics[width=\textwidth]{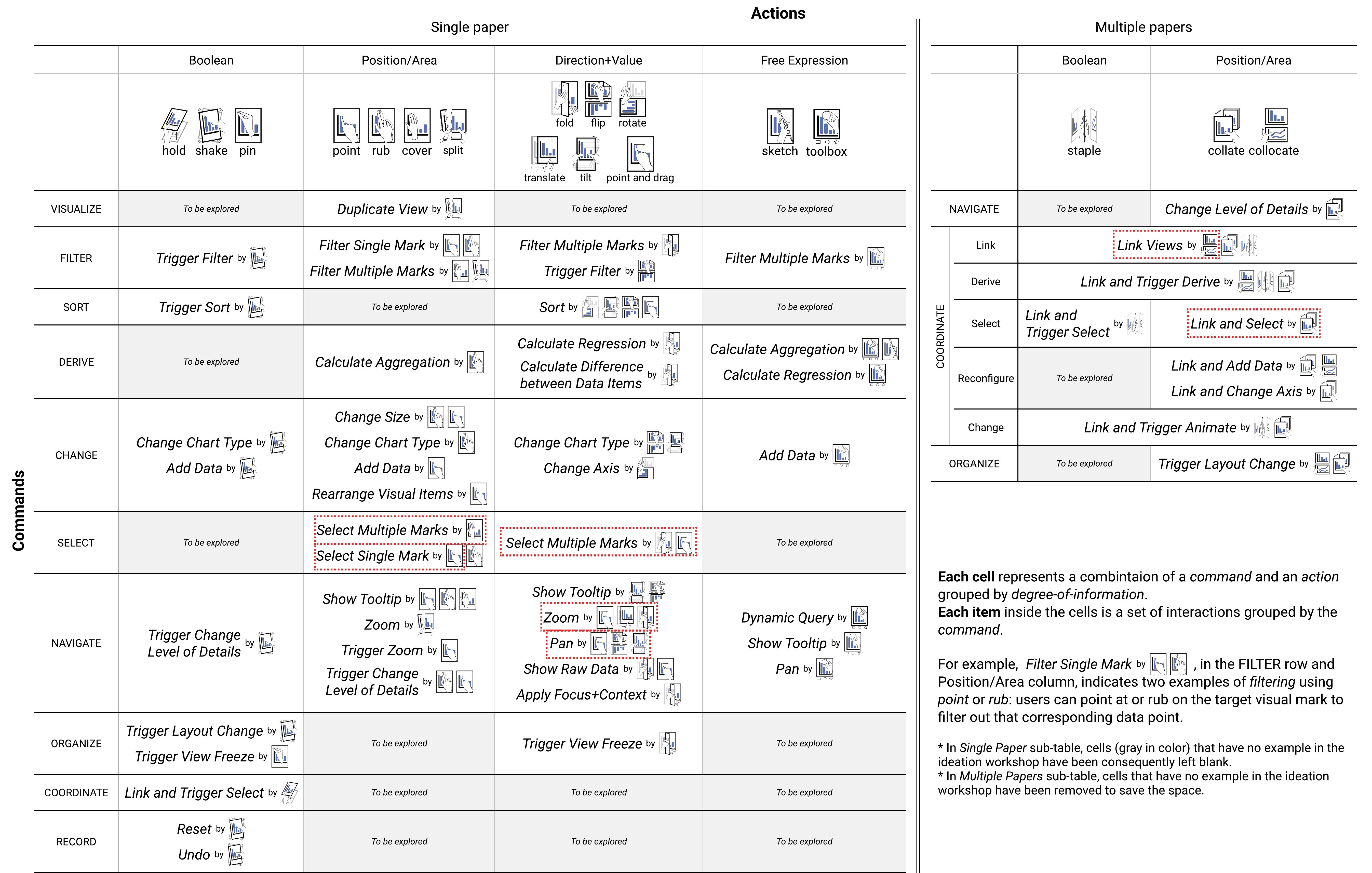}
 \caption{Interactions are summarized using the proposed design space. Sub-tables show \IntOp{}s involving a single paper (left) and multiple papers (right): \pIntFull{}s grouped by information provided (horizontally) and \opFull{}s (vertically). In each cell, we presented the \IntOp{}s found in the workshop.
 Interactions highlighted using red dashed rectangles are what we have implemented for the user study.
 }
 \label{fig:design_example}
%  \vspace{-2mm}
\end{figure*}
\section{Design Space}
\label{sec:technique}

By analyzing the \IntOp{}s resulting from the ideation workshop, we constructed a design space to facilitate the organization and creation of \pIntFull{}s for data exploration in the future.
The design space contains three dimensions: \dimOne{}, \dimTwo{}, and \dimThree{}.

\subsection{Dimension I: Commands}
\label{ssec:ops}
Dimension I, \dimOne{}, describes the low-level tasks~\cite{srinivasan2020inchorus,heer2012interactive,brehmer2013multi} \re{that an \IntOp{} achieves}{on data visualizations}{Clarify the terminology from instrumental interaction and apply it consistently (R1)}.
We listed out all \numOp{} \op{}s found in the workshop as follows.
First, participants wanted to \emph{filter} and \emph{select} data points in the visualizations.
To access more details, participants intended to \emph{navigate} (\eg, \textit{zoom}, \textit{pan}, and \textit{show tooltip}) into different charts
and \emph{derive} statistical calculations, like \texttt{mean}, \texttt{min}, and \texttt{max}.
They might also \textit{change} the chart type for different insights or update the dataset for the latest information.
For visual comparison, participants wished to \emph{sort} the data to rearrange the visual marks and \emph{organize} the visualizations in juxtaposition or superimposition.
Furthermore, participants wanted to \emph{coordinate} different charts to expand their exploration. 
For example, one participant wanted \quotestudy{the information related to this country to be highlighted in another paper (visualization) when one of them is selected.}
Lastly, participants proposed to reset the charts or undo some comments (by traversing \emph{recorded} states) to prepare another round of data exploration.

\subsection{Dimension II: Degree of Information}
\label{ssec:paperinteraction}
\setlength{\intextsep}{0pt}
\setlength{\columnsep}{3pt}

Dimension II, \dimTwo{} (DoI), 
is inspired by the notion of \emph{degree of freedom} in HCI. This dimension
describes the number of parameters an action can provide as well as the possible information. Only \pIntFull{}s that provide a matching DoI can support a given target \opFull{}.
For example, we can point at the visualization to select a data point because the point \pInt{} provides the positional information for the system to select the data point in the specified location.
However, we cannot shake the paper to select a data point because shaking cannot provide the positional information. Shaking the paper can only trigger a predefined selection.
We analyzed the DoI of each \pInt{} found in the workshop
and identified four kinds of DoI, namely,
\textit{boolean}, \textit{position/area}, \textit{direction+value}, and \textit{free expression}.
We then used these four kinds of DoI to organize the \numInt{} \pInt{}s found in the workshop
(\pIntFull{}s with an asterisk indicate \pInt{}s not presented in previous works~\cite{holman2005paper,spindler2010tangible}):

%%%%%%%%%%%%%%%%%%%%%%%%
\para{Boolean} \pInt{}s provide a \texttt{yes/no} state.
% Four \pInt{}s provide boolean information.

\begin{wrapfigure}[2]{l}{0.025\textwidth}
  \begin{center}
    \vspace{-10pt}
    % \vspace{-4pt}
    \includegraphics[width=0.025\textwidth]{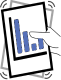}
  \end{center}
\end{wrapfigure}
\noindent
\point{Shake}: Move the paper up and down or from side to side forcefully, jerkily, and rapidly.
    \textit{Shake} provides a boolean information---whether the paper is being shook or not. For simplicity, this will involve some sort of threshold.

\begin{wrapfigure}[2]{l}{0.025\textwidth}
  \begin{center}
    \vspace{-10pt}
    % \vspace{-4pt}
    \includegraphics[width=0.025\textwidth]{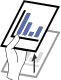}
  \end{center}
\end{wrapfigure}
\noindent
\point{Hold}: Pick up a piece of paper to the mid-air.
\textit{Hold} provides a boolean information---whether the paper is being held or lies flat on a surface.
    
\begin{wrapfigure}[2]{l}{0.025\textwidth}
  \begin{center}
    \vspace{-10pt}
    % \vspace{-4pt}
    \includegraphics[width=0.025\textwidth]{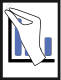}
  \end{center}
\end{wrapfigure}
\noindent
\point{Pin}*: Anchor the visualization to its current position with the pin hand gesture, similar to fixing a paper on a board using a push pin.
\textit{Pin} provides a boolean information--the paper is being pinned, or not.

\begin{wrapfigure}[2]{l}{0.025\textwidth}
  \begin{center}
    \vspace{-10pt}
    % \vspace{-4pt}
    \includegraphics[width=0.025\textwidth]{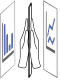}
  \end{center}
\end{wrapfigure}
\noindent
\point{Staple}:
    Place the papers face to face to mimic the metallic staple effect. \textit{Staple} provides a boolean information--whether papers are being stapled together or not.

%%%%%%%%%%%%%%%%%%%%%%%%
\para{Position/Area} \pInt{}s provide the \texttt{x}, \texttt{y}, \texttt{z} value and possibly an \texttt{area}. 

\begin{wrapfigure}[2]{l}{0.025\textwidth}
  \begin{center}
    \vspace{-10pt}
    % \vspace{-4pt}
    \includegraphics[width=0.025\textwidth]{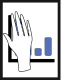}
  \end{center}
\end{wrapfigure}
\noindent
\point{Cover}*:
    Put a hand on the paper to block part of the view.
    Cover provides the position and the area covered by the hand.

\begin{wrapfigure}[2]{l}{0.025\textwidth}
  \begin{center}
    \vspace{-10pt}
    % \vspace{-4pt}
    \includegraphics[width=0.025\textwidth]{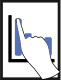}
  \end{center}
\end{wrapfigure}
\noindent
\point{Point}:
    Use a finger to point on the paper.
    \textit{Point} provides the \emph{x}, \emph{y} coordinate of the intended position of the visualization.

\begin{wrapfigure}[2]{l}{0.025\textwidth}
  \begin{center}
    \vspace{-10pt}
    % \vspace{-4pt}
    \includegraphics[width=0.025\textwidth]{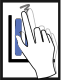}
  \end{center}
\end{wrapfigure}
\noindent
\point{Rub}:
    Point on the paper and move the finger back and forth on the paper quickly and repeatedly. 
    % Similar to pointing to the paper, but it gives a stronger tactile feeling due to the friction.
    Similar to point, \textit{rub} provides the \emph{x}, \emph{y} coordinate of the intended position of the visualization.
    
\begin{wrapfigure}[2]{l}{0.025\textwidth}
  \begin{center}
    \vspace{-10pt}
    % \vspace{-4pt}
    \includegraphics[width=0.025\textwidth]{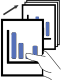}
  \end{center}
\end{wrapfigure}
\noindent
\point{Collate}:
    Stack multiple papers together.
    \textit{Collate} provides the relative position of the upper visualization to the bottom visualization.

\begin{wrapfigure}[2]{l}{0.025\textwidth}
  \begin{center}
    \vspace{-10pt}
    % \vspace{-4pt}
    \includegraphics[width=0.025\textwidth]{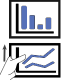}
  \end{center}
\end{wrapfigure}
\noindent
\point{Collocate}:
    Organize multiple pieces of paper side-by-side.
    \textit{Collocate} provides the relative positions of other papers.

%%%%%%%%%%%%%%%%%%%%%%%%
\para{Direction+value} \pInt{}s provide a direction and a value.
% Seven \pInt{}s are in this category: flip, tilt, rotate, fold (bend), translate, split (tear/cut), and point\&drag.

\begin{wrapfigure}[2]{l}{0.025\textwidth}
  \begin{center}
    \vspace{-10pt}
    % \vspace{-4pt}
    \includegraphics[width=0.025\textwidth]{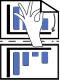}
  \end{center}
\end{wrapfigure}
\noindent
  \point{Flip}:
    Turn the paper's front side back or the back side front.
    Flipping along different edges of the paper provides the direction information and the current state of the paper---facing up or down.
    
\begin{wrapfigure}[2]{l}{0.025\textwidth}
  \begin{center}
    \vspace{-10pt}
    % \vspace{-4pt}
    \includegraphics[width=0.025\textwidth]{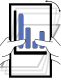}
  \end{center}
\end{wrapfigure}
\noindent 
\point{Tilt}:
    Slant the view plane to a different angle than its normal viewing position.
    Tilting vertically and horizontally provides different direction information, and the tilt angle provides the value.
    
\begin{wrapfigure}[2]{l}{0.025\textwidth}
  \begin{center}
    \vspace{-10pt}
    % \vspace{-4pt}
    \includegraphics[width=0.025\textwidth]{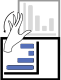}
  \end{center}
\end{wrapfigure}
\noindent
\point{Rotate}:
    Reorient the paper to a different angle.
    Similar to tilt, \textit{rotate} provides the direction of rotation and the rotation degree.

\begin{wrapfigure}[2]{l}{0.025\textwidth}
  \begin{center}
    \vspace{-10pt}
    % \vspace{-4pt}
    \includegraphics[width=0.025\textwidth]{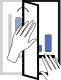}
  \end{center}
\end{wrapfigure}
\noindent
\point{Fold (bend)*}:
    Fold or bend a piece of paper over to cover other parts of itself.
    \textit{Fold/bend} provides the folding/bending direction and the portion of the cover.
    
\begin{wrapfigure}[2]{l}{0.025\textwidth}
  \begin{center}
    \vspace{-10pt}
    % \vspace{-4pt}
    \includegraphics[width=0.025\textwidth]{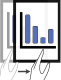}
  \end{center}
\end{wrapfigure}
\noindent
\point{Translate}:
    Move the paper up and down, left and right, also close or far from the eyes.
    \textit{Translate} provides the direction and magnitude of the movement.

\begin{wrapfigure}[2]{l}{0.025\textwidth}
  \begin{center}
    \vspace{-10pt}
    % \vspace{-4pt}
    \includegraphics[width=0.025\textwidth]{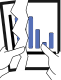}
  \end{center}
\end{wrapfigure}
\noindent
 \point{Split (Tear/Cut)*}:
    Tear or cut the paper into two parts, splitting up the content.
    \textit{Tear/Cut} provides the tearing/cutting direction and the size of the resulting parts.

\begin{wrapfigure}[2]{l}{0.025\textwidth}
  \begin{center}
    \vspace{-10pt}
    % \vspace{-4pt}
    \includegraphics[width=0.025\textwidth]{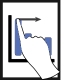}
  \end{center}
\end{wrapfigure}
\noindent
\point{Point\&drag}:
    Point and drag one or multiple fingers on the paper, such as drag and pinch gestures.
    \textit{Point and drag} utilizes time to create the direction and moving distance.

%%%%%%%%%%%%%%%%%%%%%%%%
\para{Free expression} \pIntFull{}s 
% are the most expressive.
% These \pInt{}s 
can provide an expression beyond numerical values.
% Toolbox and sketch are included in this category.

\begin{wrapfigure}[2]{l}{0.025\textwidth}
  \begin{center}
    \vspace{-10pt}
    % \vspace{-4pt}
    \includegraphics[width=0.025\textwidth]{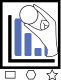}
  \end{center}
\end{wrapfigure}
\noindent
\point{Toolbox}:
    Utilize other papers with different shapes, colors, and text annotations as interactive widgets (\eg, buttons, menus, and sliders) for user input.
    Depending on the design of the paper widget, a toolbox can provide any expression to manipulate the visualization.

\begin{wrapfigure}[2]{l}{0.025\textwidth}
  \begin{center}
    \vspace{-10pt}
    % \vspace{-4pt}
    \includegraphics[width=0.025\textwidth]{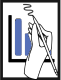}
  \end{center}
\end{wrapfigure}
\noindent
\point{Sketch}:
    Use a pen or digital pen to write or draw on the paper.
    Depending on the predefined \op{}s, free-form sketching or writing can provide any expression to interact with the visualization.

Note that this analysis is capturing only those mappings discussed in the workshop. For example, there could potentially be a multitude of ways to shake a paper, \eg, shake vertically, shake horizontally, shake multiple times. Our design space aims at a first overview of possible and feasible \IntOp{}s and thus these variations are not considered.

\subsection{Dimension III: Number of Paper Sheets Involved}
Dimension III, \dimThree{}, describes the number of papers involved in the \pIntFull{}. 
Paper actions involving one paper target at single view manipulation, while \pIntFull{}s involving multiple papers supports multiview manipulation and analysis.

\para{Single paper.} There are 15 \pInt{}s (as shown in \autoref{fig:design_example}) found to involve one piece of paper in the workshop.
For example, participants pointed at one paper and folded one paper.

\para{Multiple papers.} Three \pInt{}s (\ie, collate, collocate, and staple) were found to involve two or more pieces of paper.
These \pInt{}s allow users to organize multiple sheets of paper into different layouts or use visualization as an object to interact with other visualizations.
% \re{}{People can manipulate two papers easily, while it is harder for more papers. They might stacked multiple paper sheets for one hand manipulation.}{I wonder if there is utility in distinguishing between one, two, and multiple papers in Section 4.3 (Dimension III) as people can typically manipulate two papers at the same time with ease, but more than two is harder. (R2)} 

\subsection{Supporting Commands through Paper Actions}

% \begin{figure*}[t]
%  \centering % avoid the use of \begin{center}...\end{center} and use \centering instead (more compact)
%  \includegraphics[width=\textwidth]{figures/final design space.pdf}
%  \caption{Interactions summarized using the proposed design space. Sub-tables show \IntOp{}s involving a single paper (left) and multiple papers (right): \pIntFull{}s grouped by information provided (horizontally) and \opFull{}s (vertically). In each cell, we presented the \IntOp{}s found in the workshop.
%  Interactions that are highlighted using red dash rectangles are what we have implemented for the user study.
%  }
%  \label{fig:design_example}
% %  \vspace{-2mm}
% \end{figure*}

\begin{figure}[t]
 \centering % avoid the use of \begin{center}...\end{center} and use \centering instead (more compact)
 \includegraphics[width=0.9\linewidth]{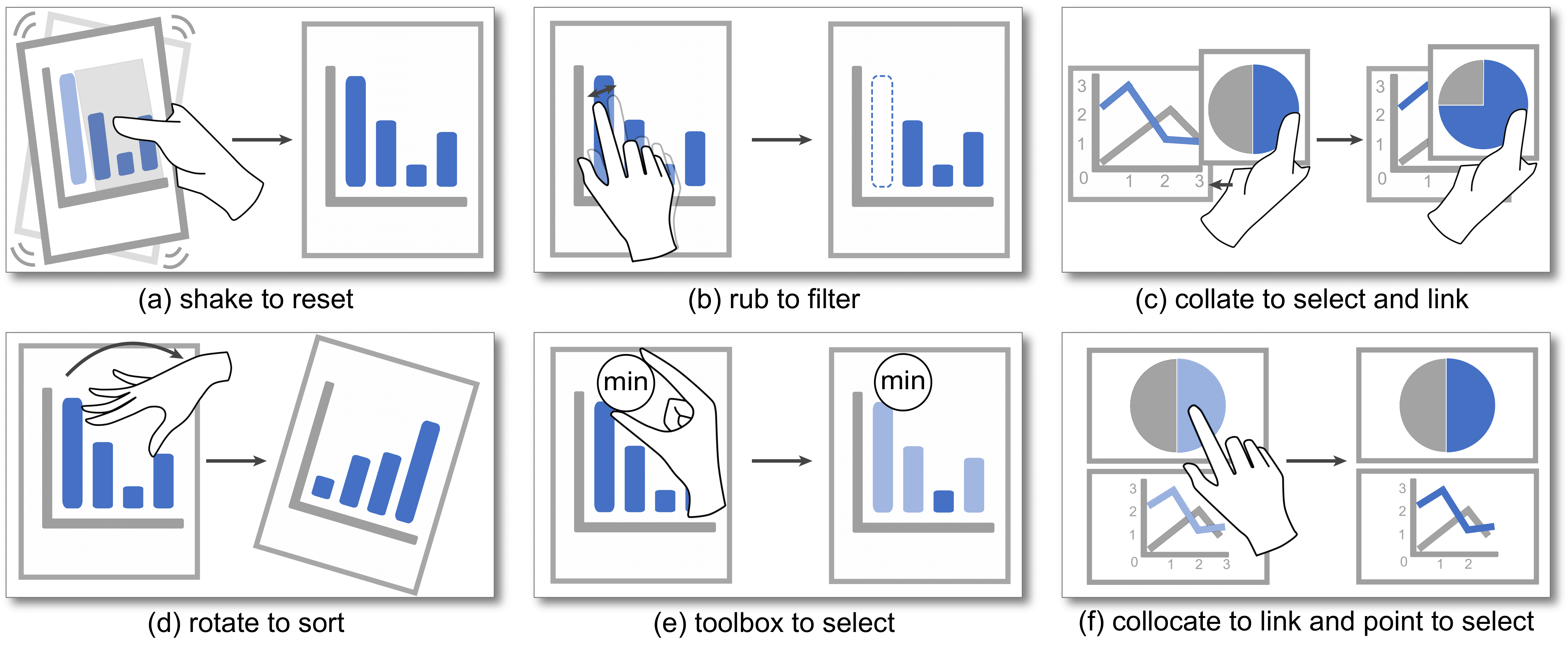}
 \caption{Illustrations of \IntOp{} examples. (a) - (e) cover examples in each DoI category. (a) shows a boolean \IntOp{}, (b) and (c) show a position \IntOp{} and a multiview position \IntOp{}, (d) shows a direction+value \IntOp{} , and (e) shows a free expression \IntOp{}. (f) shows a combination of two \IntOp{}s to link and select.
 }
 \label{fig:int_ops_illustration}
%  \vspace{-2mm}
\end{figure}

% In the remaining part of the paper, we use the notion of \mapping{$\pInt{}$}{$\op{}$} (\eg, \mapping{point}{select single mark}) to describe an \IntOp{}: using a \pIntFull{} $I$ to support an \opFull{} $O$.

Based on the design space, % formed,
% Based on the design space formed by the above three dimensions,
% \numInt{} \pIntFull{}s and \numOp{} \opFull{}s,
we describe the collected \IntOp{}s from the workshop as shown in \autoref{fig:design_example}.
The figure shows \IntOp{}s involving one paper sheet (left) and multiple paper sheets (right). In each sub-table, we have DoI (with the corresponding \pIntFull{}s) listed horizontally and \opFull{}s listed vertically.
% Each cell in the design space represents 
% how one kind of degrees-of-information supports a \opFull{}.
Each item in a cell represents one or more \IntOp{}s grouped by the \op{}s. 
% The number on the right of each \pIntFull{} icon represented the number of participants having the same idea.
For example, 
\smash{\raisebox{-3pt}{\includegraphics[height=12pt]{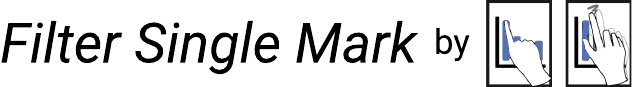}}}
indicates two \IntOp{}s: \mapping{point}{filter-single-mark}, and \mapping{rub}{filter-single-mark}.
Other combinations (\eg, \mapping{translate}{select-single-mark}) that had no practical solutions proposed in the workshop, we have left blank.
% As mentioned in Sec.~\ref{ssec:ops}, 
Below, we explain the details of the design space organized by DoI.

    \para{Boolean \IntOp{}s} (14/\numComb{}) are mainly used as a trigger to activate or deactivate \op{}s.
    For example, it could be used as \mapping{shake}{trigger-filter} (\smash{\raisebox{-3pt}{\includegraphics[height=12pt]{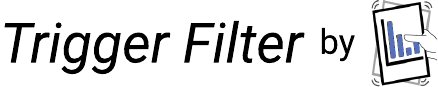}}}) or \mapping{shake}{reset} (\autoref{fig:int_ops_illustration}(a)).
    Moreover, we can trigger \op{}s (\eg, derive) with multiview by stapling paper sheets. 
    The number of Boolean \IntOp{}s is low, probably because
    the expressiveness (ability to convey users' intentions to the visualization) of these \IntOp{}s is low.
    % users grant the least control to the visualization.

    \para{Position/Area \IntOp{}s} (33/\numComb{})
    % are relatively higher than Boolean \IntOp{}s. 
    allow users to directly communicate with specific visualization components of the visualization
    since it provides position data for the system to locate visual elements, \ie, \textit{x} and \textit{y} location.
    % These \IntOp{}s give more control for users to manipulate specific visualization components of the visualization.
    In addition to triggering the filter \op{} in the Boolean \IntOp{},
    users can now specify the visual mark to be filtered out by pointing or rubbing (\smash{\raisebox{-3pt}{\includegraphics[height=12pt]{figures/Design Space Example/filter_single_mark.pdf}}}).
    Figure~\ref{fig:int_ops_illustration}(b) illustrates that users filter a bar on a printed visualization using the rubbing gesture.
    % users generated more ideas and techniques in this area.
    % For examples, six \IntOp{} can be summarized by 17 ideas for selection \op{}. 
    % By providing the position or area information on the visualization, different \op{}s can be done, such as, 1) select mark or legend item on the visualization by pointing, 2) filtering marks by rubbing, and 3) reconfiguring
    Moreover, by involving multiple papers, multiple visualizations can be coordinated using their spatial relationship for more complex multivariate data exploration.
    There are 12/33 \IntOp{}s involving multiple papers to perform navigation, coordination, and organization.
    % \re{, similar to using mobile devices~\cite{langner2017vistiles}.}{link with previous work}
    For example, as shown in \autoref{fig:int_ops_illustration}(c),
    % users can reconfigure the upper visualization to show more data by stacking two visualizations
    users can overlay one paper over the other one to pick a specific timestamp from the bottom visualization and update both visualizations
    (\smash{\raisebox{-3pt}{\includegraphics[height=12pt]{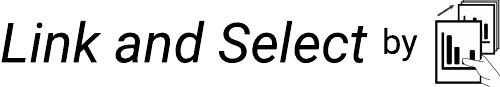}}}).
    
    \para{Direction+Value \IntOp{}s} (25/\numComb{}) provide more information than the position.
    They allow users to perform \op{}s that require directional information such as sorting in ascending or descending order by rotating, tilting, flipping, or dragging (\smash{\raisebox{-3pt}{\includegraphics[height=12pt]{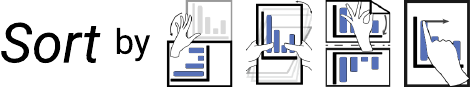}}}) as shown in \autoref{fig:int_ops_illustration}(d).
    Moreover, direction+value \IntOp{}s can also be transformed into area information.
    For example, users can fold the paper in \emph{x} or \emph{y} direction to select or filter an area of visual marks covered by the folded part of the paper.
    Some of the interactions, such as \mapping{tilt}{pan} and \mapping{translate}{zoom}, have previously been proposed for mobile devices~\cite{spindler2014pinch} or tabletop paper lens~\cite{spindler2010tangible}.
    
    \para{Free expression \IntOp{}s} (8/\numComb{}) can deal with more complex \op{} that are derived and encoded by utilizing extra objects, \ie, pen or customized toolbox.
    With their expressive power, these \IntOp{}s can support advanced filtering, querying, or calculation, such as directly picking the elements with the minimum value using a circle shaped representing a \texttt{min} \op{} (\smash{\raisebox{-3pt}{\includegraphics[height=12pt]{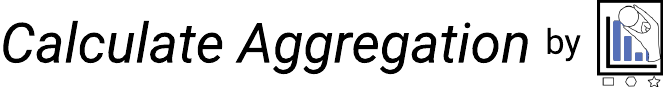}}}, \autoref{fig:int_ops_illustration}(e)).
    % \re{For example, papers can be folded or cut into different shapes, such as rectangular or circular, which could be used as zoom and round function respectively~\cite{spindler2010tangible}\ben{<---This is not clear to me what you mean here}.}{link to past work}
    % \newline
    % However, the number of \IntOp{}s collected for this category is lower than that of Position/Area and Direction+Value,
    % probably because that it is hard to think of utilizing extra objects due to the more complicated usage and generalizability issue.

\vspace{1mm}
Overall, our design space provides designers with a structured way to design paper interactions on printed visualizations.
% The designed \IntOp{}s can be used together as building blocks for more complex tasks due to the use of \dimOne{}.
% For example, we can have \mapping{point}{select single mark} and \mapping{collocate}{link} working together for multiview analysis (as illustrated in \autoref{fig:int_ops_illustration}(f)).
It provides an overview over the feasibility of the \IntOp{}s.
% by using \dimTwo{}.
For example, we cannot have \mapping{shake}{select-single-mark} as the \textit{shake} action only provides a Boolean input to the selection command. 
Designers can look up Position/Area \IntOp{}s to choose an action for the selection command.
\section{User Study}
\label{sec:eval}

\begin{table*}
\small
\centering
\caption{Interactions implemented in this study with their names and a brief description.}
\label{tab:techniques}

\begin{tabular}{l|l|l} 
\toprule
% \multicolumn{1}{l|}{\textbf{Commands}} 
\textbf{Commands} & \textbf{Actions} & \textbf{Description} \\ 

\hline
\multirow{3}{*}{Select-an-interval} & Point\&Drag & \begin{tabular}[c]{@{}l@{}}Swipe the finger on the paper sheet\\$\implies$ select the data points in the range of the axis brushed\end{tabular} \\ 
\cline{2-3}
 & Cover & \begin{tabular}[c]{@{}l@{}}Cover part of the visualization with flat hand\\$\implies$ select the data points NOT in the range of the axis covered\end{tabular} \\ 
\cline{2-3}
 & Fold & \begin{tabular}[c]{@{}l@{}}Fold the paper to cover a large (or small) portion of the axis\\$\implies$ select the data points NOT in the range of the axis covered\end{tabular} \\ 
\hline
\multirow{3}{*}{Zoom} & Pinch & \begin{tabular}[c]{@{}l@{}}Two-finger pinch outward (or inward) on the visualization\\$\implies$ zoom-in (or zoom-out) of the visualization\end{tabular} \\ 
\cline{2-3}
 & Translate & \begin{tabular}[c]{@{}l@{}}Move the paper closer to (or farther from) the camera\\$\implies$ zoom-in (or zoom-out) of the visualization\end{tabular} \\ 
\cline{2-3}
 & Fold & \begin{tabular}[c]{@{}l@{}}Fold the paper to cover a large (or small) portion of the axis\\$\implies$ zoom-in (or zoom-out) to the portion of visualization NOT covered\end{tabular} \\ 
\hline
\multirow{3}{*}{Pan} & Point\&Drag & \begin{tabular}[c]{@{}l@{}}Finger scroll left (right, up, or down) on the visualization\\$\implies$ pan the visualization rightward (leftward, downward, or upward)\end{tabular} \\ 
\cline{2-3}
 & Tilt & \begin{tabular}[c]{@{}l@{}}Slant the paper to the left (right, up, or down) relative to the ground\\$\implies$ pan the visualization rightward (leftward, downward, or upward)\end{tabular} \\ 
\cline{2-3}
 & Flip & \begin{tabular}[c]{@{}l@{}}Flip the paper from left to right (right to left, up to down, or down to up)\\$\implies$ pan the visualization leftward (rightward, upward, or downward)\end{tabular} \\ 
\hline
\multirow{2}{*}{Link\&Select} & Collate & \begin{tabular}[c]{@{}l@{}}Put one visualization on top of another and center it to specific position relative to the bottom one\\$\implies$ connect the two visualizations, select the data point from the bottom one, and update both visualizations\end{tabular} \\ 
\cline{2-3}
 & Collocate\&Point & \begin{tabular}[c]{@{}l@{}}Put two visualizations side by side and point at the visualization\\$\implies$connect the two visualizations, select the data point, and update both visualizations\end{tabular} \\
\bottomrule
\end{tabular}
\end{table*}

\re{We conducted a controlled user study to 
\emph{(G-Preferences)} gather and reason on user preferences and to \emph{(G-Viability)} investigate the practical viability of the \IntOp{}s.}{Our design space helps designers to design feasible paper interactions for data visualization. Apart from the design insights, we want to further investigate paper interactions' functionality in real practice. Thus, we conducted a controlled user study with a proof-of-concept prototype to investigate user preferences \emph{(G-Preferences)} and practical viability \emph{(G-Viability)} of the \IntOp{}s.}{Motivate the study in relation to the design space}

\subsection{Prototype}
\begin{figure}[tb]
 \centering % avoid the use of \begin{center}...\end{center} and use \centering instead (more compact)
 \includegraphics[width=\linewidth]{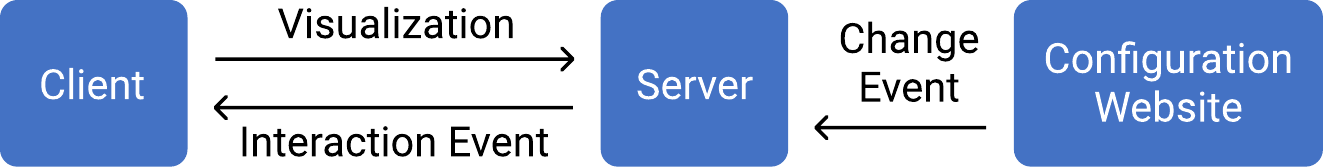}
 \caption{This figure shows the framework of the prototype and relationship between client, server, and configuration website.}
 \label{fig:framework}
\end{figure}

\begin{figure*}[t]
 \centering % avoid the use of \begin{center}...\end{center} and use \centering instead (more compact)
 \includegraphics[width=\linewidth]{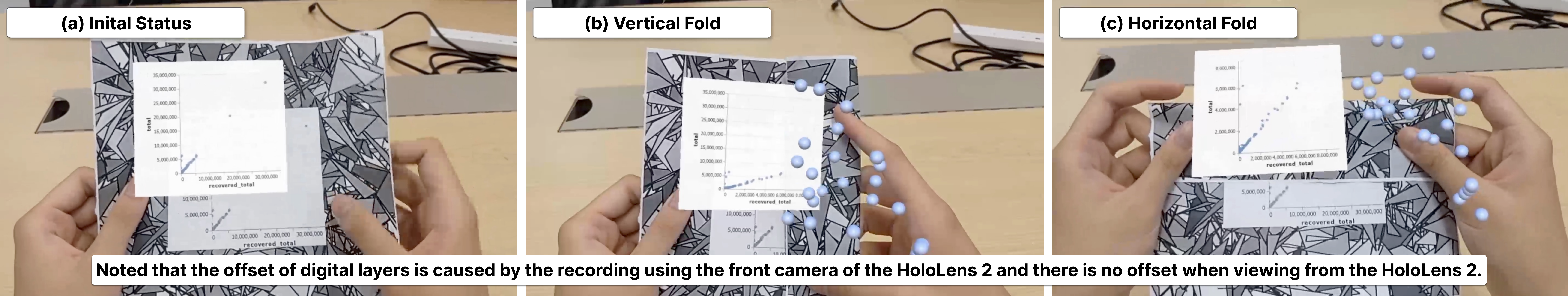}
 \caption{This figure demonstrates an example of \mapping{fold}{zoom}: (a) the initial scatter plot and its changes after (b) vertically and (c) horizontally folding the paper. The right hand joints are visualized using the white spheres.}
%  \vspace{-4mm}
 \label{fig:realfootage}
\end{figure*}

We simplify our study by focusing on those \op{}s used for view manipulation as described by Heer and Shneiderman~\cite{heer2012interactive}.
We implemented the 11 \IntOp{}s listed in Table~\ref{tab:techniques}. An example of using \mapping{fold}{zoom} can be found in \autoref{fig:realfootage}.
% \ben{The table header is not good. 1) what is the first column showing? is that necessary? The second column is shoiwng commands, right? the 3rd one is showing actions, right? Name those coloumns accordingly. I don't don't want to mess with your table layout.} 
We built the experimental prototype using a client-server model with a WebSocket for the network communication as shown in \autoref{fig:framework}.

\begin{figure}[t]
 \centering % avoid the use of \begin{center}...\end{center} and use \centering instead (more compact)
 \includegraphics[width=0.75\linewidth]{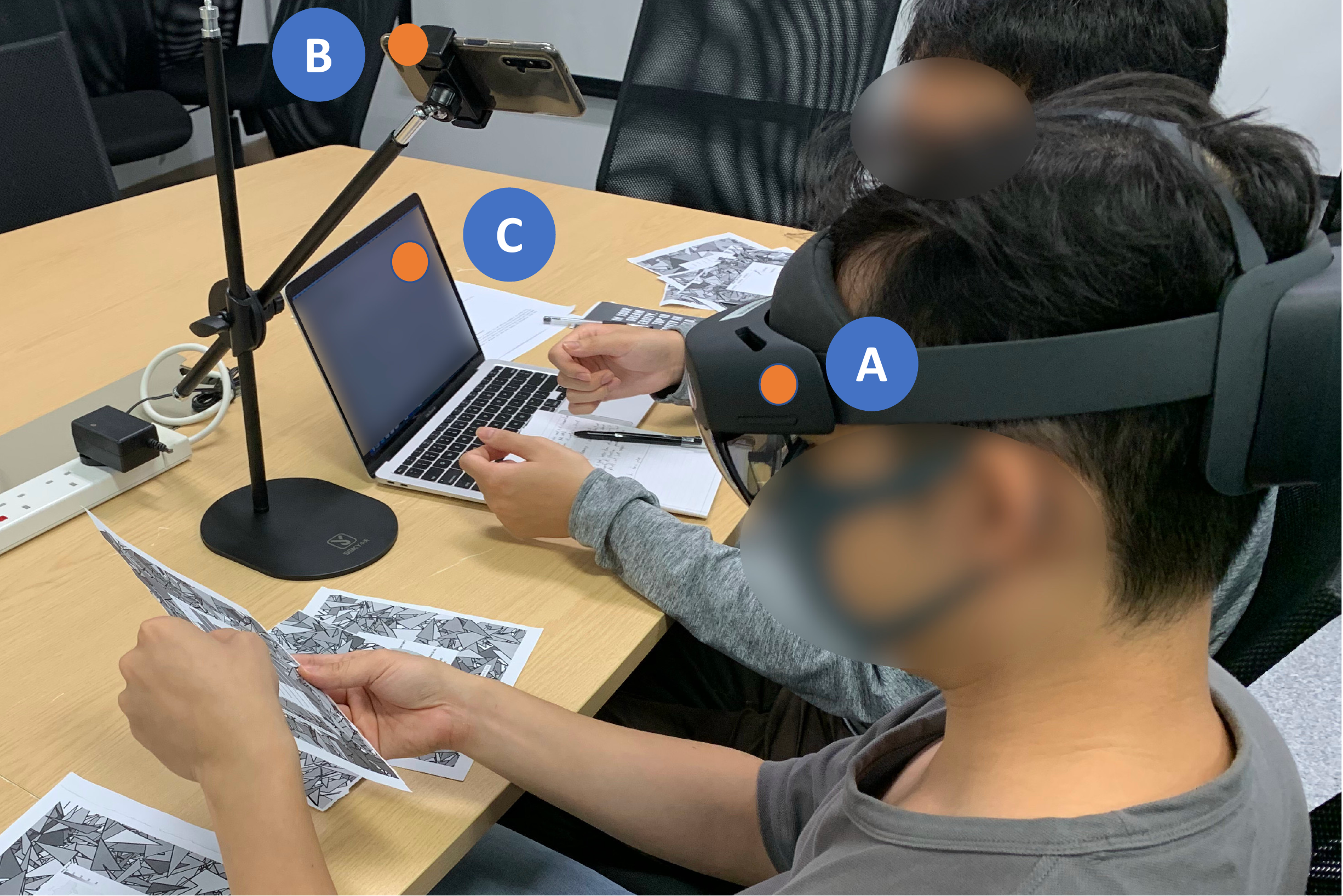}
 \caption{This figure shows the experiment setting: (a) the participant were wearing the HoloLens 2 to view the digital visualization in AR; (b) both of the participant's hands were recorded by both a smartphone on a phone stand; (c) a laptop was used by one of the authors to change the settings of the current activated \IntOp{}s.}
 \vspace{-4mm}
 \label{fig:setting}
\end{figure}

\begin{itemize}[leftmargin=*]
    \item The \textbf{client} app runs on the Hololens 2 to detect users' hand gestures and papers' status.
    The newer HoloLens 2 provides a diagonal $52^{\circ}$ field of view and two-handed fully articulated hand tracking.
    \re{The pose of paper sheets and users' hands are obtained through Vuforia\footnote{https://developer.vuforia.com/} image target detection and Microsoft's Mixed Reality Toolkit\footnote{https://docs.microsoft.com/en-us/windows/mixed-reality/mrtk-unity/?view=mrtkunity-2021-05} on HoloLens 2 respectively, as shown in \autoref{fig:setting}(a).}{Hand-pose data was collected through Microsoft’s Mixed Reality Toolkit
    % \footnote{https://docs.microsoft.com/en-us/windows/mixed-reality/mrtk-unity/?view=mrtkunity-2021-05}
    on HoloLens 2. For pose tracking of printed visualizations, we used Vuforia
    % \footnote{https://developer.vuforia.com/}
     image tracking. Rich feature patterns were added on all sides of the paper for more accurate tracking.}{Clarify technical infrastructure (R4)}
    % \ben{can you point to a figure here? Is it Fig 6?)}
    Depending on the relative position and direction of fingers and paper sheets, different \pIntFull{}s and gestures are recognized as action events (\autoref{tab:techniques}). \re{}{For example, when two sheets were placed close together, a collocate action event was created (\autoref{fig:teaser}(e)).}{Same as above}
    % \ben{where does this triggering happen? On the client app? or on the server?}
    % After observing  the \pIntFull{}, the client app 
    After observing the events, the client app sends these events to the server.

    \item The \textbf{server} receives the \pIntFull{} events with their parameters (\eg, positions in the case of a Position/Area action) and updates the visualization through a corresponding command.
    We maintain a Vega~\cite{satyanarayan2015reactive} specification for each visualization on the server. The updated visualization is returned to the client.

    \item The \textbf{configuration website} with dropdown menus allows the conductor to change the selected \IntOp{}s based on the task and participants' needs (\autoref{fig:setting}(c)).
    The configuration is updated to the server and the effect is immediately reflected in the client app.
\end{itemize}

% \begin{figure}[t]
%  \centering % avoid the use of \begin{center}...\end{center} and use \centering instead (more compact)
%  \includegraphics[width=\linewidth]{figures/examples_seen_by_users.pdf}
%  \caption{This figure shows screenshots of four \IntOp{}s seen by users via HoloLens 2. Noted that the offset is caused by the recording and there is no offset viewing from the HoloLens 2.}
%  \label{fig:realfootage}
% \end{figure}

\begin{figure*}[tb]
 \centering % avoid the use of \begin{center}...\end{center} and use \centering instead (more compact)
 \includegraphics[width=\textwidth]{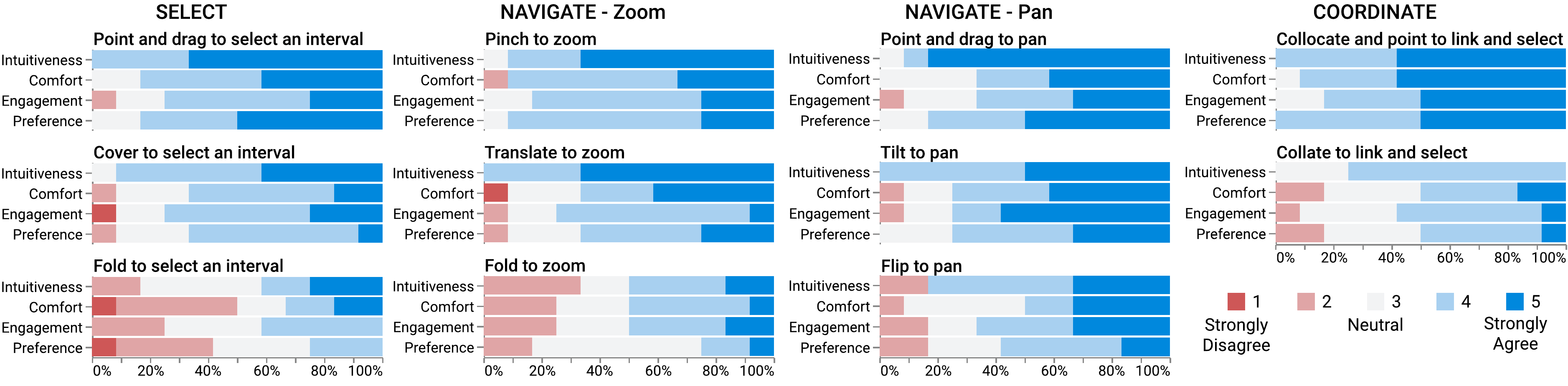}
 \caption{This figure shows participants' ratings on ``intuitiveness'', ``comfort'', ``engagement'', and ``overall preference'' for each \IntOp{}s in task 1.
%  Ranging from Strongly Disagree (1) to Strongly Agree (5).
%  for ``intuitiveness'', ``comfort'', ``engagement'' and ``overall preference''
 }
%  \vspace{-4mm}
 \label{fig:result}
\end{figure*}

\subsection{Setup and Participants}
We recruited 12 university students (P1-P12; aged between 22 and 30; 6 males and 6 females).
None of them had participated in the ideation workshop.
The distribution of their visual analysis experience was ``none'' (3), ``novice'' (5), ``knowledgeable'' (4), and ``expert'' (0).
The distribution of their AR experience was ``none'' (2), ``novice'' (9), ``knowledgeable'' (1), and ``expert'' (0).
The distribution of their daily paper usage was ``0 day per week'' (1), ``1-2 days per week'' (2), ``3-6 days per week'' (1), and ``every day'' (8). 
Overall, participants are mainly novices in both visual analysis and AR, and daily paper users.
% The participants were equipped with the HoloLens 2.
% This model was selected due to its egocentric vision, mobility, and precise 3D hand tracking.
% a mobile phone with a 6.26-inch screen and a resolution of $2340\times1080$ held by a phone holder for the whole study, as shown in.
% Each participant received \$13 to \$17 as compensation, depending on the duration.
% The study conductor took notes on the participants' comments during the study. Moreover, for reference after the study, all sessions were audio and video recorded using a mobile phone (as shown in~\autoref{fig:setting}), as well as a low-resolution screen capture (\ie, a 428x240p 15fps 0.6Mbit stream) on the HoloLens 2.
All sessions were
% audio and video
recorded using a mobile phone, as shown in \autoref{fig:setting}(b).

\subsection{Procedure}
The study consisted of an introduction, 
two tasks, and a semi-structured interview.
Each participant received \$13-\$17 as compensation according to a 90-120 minute study time.
% It lasted 1.5-2 hours.
% Before the introduction, we asked participants to fill out a questionnaire on their backgrounds. 

\para{Introduction ($\sim$10 mins).}
We first introduced the study background and procedure, and then asked participants to sign the consent form.
After that, participants were asked to put on the AR HMD and adjust the device until they felt comfortable and could see the AR content attached to the printed visualization clearly.

% task 1
\para{Task 1: Unit Evaluation ($\sim$60 mins).}
% To build familiarity and evaluate each individual \IntOp{},
To assess \emph{G-Preferences} for each \IntOp{},
we asked participants to perform all 11 \IntOp{}s on a set of printed visualizations showing the latest Covid-19 dataset (\eg, a scatter plot with total confirmed cases against total recovered cases, \autoref{fig:realfootage})
% \ben{say exactly what had been shown. E.g., A single linechart with case numbers? show a figure if possible}
, the source of which was the same as the workshop.
We counterbalanced the sequence of \opFull{}s and also the sequence of \pInt{}s per \op{} using the balanced Latin Square method~\cite{bradley1958complete}.
For each \op{}, we first introduced the task to the participants and then the participants started to try each \IntOp{} with the procedure below:
\begin{enumerate}[leftmargin=*,itemsep=0pt,topsep=0pt,partopsep=0pt, parsep=0pt]
    \item The study conductor demonstrated the \IntOp{} to the participant.
    \item The participant performed the \IntOp{} five or more times
    successfully.
    \item The participant rated the \IntOp{} on metrics widely adopted in previous research on interactions in AR, \ie, intuitiveness, comfort, engagement, and overall preference~\cite{samini2017popular}, by filling in a questionnaire.
    (To get ratings independently and not implementation specific, participants were told that it was not necessary to compare it with other presented \IntOp{}s, and it was assumed that the HoloLens 2 worked without technical issues~\cite{gupta2020replicate}.)
    \item A series of follow-up questions were asked to obtain further comments and in-depth rationales for the ratings.
\end{enumerate}
% Participants were encouraged to think aloud while trying each \IntOp{}.
% It took the participants around one hour to experience all of the \IntOp{}s and express their feelings and comments.
After Task 1, participants were asked if they had any discomfort. A five-minute break was given based on the participants' needs.

% task 2
\para{Task 2: Free-Form Exploration ($\sim$15 mins).}
% In Task 2, we wanted to see how \IntOp{}s are used together in practice by the participants.
For \emph{G-Viability} (whether people can use and how they use \IntOp{}s), we asked participants to use the above-mentioned \IntOp{}s to answer a question and explore the data freely within the given 15 minutes time frame.
We introduced a set of five visualizations (two visualizations are shown in \autoref{fig:teaser}) on a different dataset from Task 1 (\ie, worldwide university rankings in 2016) to reduce the effect of memorizing the dataset from Task 1.
% \ben{again, point to a small figure showing what the user saw. There are many places in the paper where we can text if we get over the page limit.}.
% To examine the trade-off and combination of different \IntOp{}s, we set the constraint that one operation could be assigned with only one interaction.
To initialize a set of \IntOp{}s for participants to perform the Task 2 as the initial setting for free exploration, we picked those \IntOp{}s with the highest preference for each \op{} from Task 1.
% \ben{from Task1? I don't understand what you need to initialize them with? Wasn't this free-form exploration}.
For \IntOp{}s with the same preference rating, we let the participants choose the \IntOp{}.
% We first assigned the \IntOp{} with the highest overall performance as rated in Task 1 to the corresponding operation at the beginning. For interaction with the same highest scores, we let the participants to choose the interactions.
In addition, participants were encouraged to tell us when they want to change and explore different \IntOp{} mappings, \eg, changing \mapping{pinch}{zoom} to \mapping{translate}{zoom}), and then we change the setting for them. 
% \ben{how did they change that? was that a menu? or did they tell you?} 
To kick-start exploration, we asked participants to answer the following question: ``\textit{Given the line chart, what is the trend of MIT’s total score from 2011 to 2016?}''. Participants were required to zoom (select, and pan if necessary) since the line chart was complex and cluttered at times, as shown \autoref{fig:teaser}(b).
% , which were inspired by the task setting in Srinivasan~\etal's work~\cite{srinivasan2020inchorus}.
After answering this question, participants could use the remaining time to explore on the five visualizations freely.
During free exploration, participants were asked to think aloud about what they were doing
and how they planned to perform an \IntOp{}.
% We also encouraged participants to express their feelings when performing the \IntOp{}s since
% we were interested in the participants' behaviours and feedback while performing data exploration with multiple \IntOp{}s.
% To understand the comfort of interacting with paper sheets for data exploration,
% we asked the participants whether they had any discomfort again after Task 2.

% debriefing
\para{Post-Study Interview ($\sim$15 mins).}
% To further understand how participants used the \IntOp{}s in the study 
% and their willingness to use \pIntFull{}s to explore visualizations in real-life scenarios, we have conducted a semi-structured interview after completing all of the tasks.
% We asked questions about their comments on the tasks---for example, \textit{``how did you solve interaction conflicts, if any?"}---and asked open questions towards the interaction techniques.
% Moreover, we asked follow-up questions according to the participant's performance in the study.
% This post-study interview lasted approximately 10 minutes.
We conducted a semi-structured interview with nine questions in three topics: preference, usefulness, and possible new \IntOp{}, at the end of the study.

% interaction-based questions:
% What techniques do you still remember?
% How do you decide to choose interaction?

% usability and usage scenarios of the \pIntFull{} interface questions:
% Will you be willing to use this interface in the future if technically available?
% What are the usage scenarios? Can you give examples?
% Do you look forward to a complete prototype system?
% Overall, what do you think are the biggest advantages and disadvantages of this type of interaction interface for data visualization?
% Do you think papers are a good medium to interact with? Any concerns?

% new idea questions:
% Any new interaction ideas?
% Any new operations?

\subsection{Results}
\label{ssec:result}
We report on participants' quantitative ratings and verbal feedback for the \IntOp{}s from both tasks as well as the semi-structured interview in the user study.
Overally, with respect to \textit{G-Preferences}, we found that participants rated the proposed \IntOp{}s intuitive and engaging to use. With respect to \textit{G-Viability}, they enjoyed interacting with the printed visualizations using \pIntFull{}s with different affordances.

\subsubsection*{Preference and feedback for \IntOp{}s}
% \ben{use past tense when you report on experiments}
% \ben{Rename G1 'G-Prefereces' and G2 G-viability, or make both terms (preferences, viability) \textsc. That's easier to read than G1, G2..} 

\noindent Figure~\ref{fig:result} shows the ratings of 12 participants on \textit{intuitiveness}, \textit{comfort}, \textit{engagement}, and \textit{overall preference} across all \IntOp{}s for each \opFull{}. Overall, participants were positive for most of the \IntOp{}s from each dimension. 
% Below we present their detailed reasons.
Below we present the users' preferences and feedback grouped by the interaction \op{}s. The number inside the brackets indicates the median of the ratings.
% supported by the \IntOp{}s.

% {'link and brush': {'collocate to link and brush': 12, 'collate to link and brush': 4}, 'pan': {'tilt to pan': 6, 'point to pan': 9, 'flip to pan': 3}, 'select an interval': {'cover to select an interval': 4, 'point to select an interval': 9, 'fold to select an interval': 1}, 'zoom': {'translate to zoom': 5, 'point to zoom': 8, 'fold to zoom': 2}}

% \ben{In the following, you use the commands specified in fig.3. At the same time Fig 3 does not show the sub-commants (select-an-interval, etc..). Can you add these subcommands to Fig 3? E.g., 'Navigate' is not really a command, but 'pan' or 'zoom' is. } 
% \ben{make sure to convert everything to past tense!!! I did but might have missed some}
% \ben{below, I removed the individual mentions of the command since that is in the paragraph headers. It's easier to read.}
% \ben{I removed the small distribution charts from your makro. Feel free to reinclude them but that information is shown in Figure 8 and it's breaking the reading flow.} 

\para{Select-an-interval.} In this task, participants were asked to select a range of bars in a given bar chart.
% Among the three given \IntOp{}s (\ie, \mapping{point\&drag}{select an interval}, \mapping{cover}{select an interval}, and \mapping{fold}{select an interval}), 
Participants preferred the \textit{point\&drag} \pInt{} the most \statnew{4.33}{0.78}{4.5}{5}{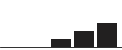}.
% Participants rated \mapping{point\&drag}{select an interval} as the most intuitive \statnew{4.67}{0.49}{5}{5}{figures/results/intuitive_point_to_select_an_interval.png}, comfortable \statnew{4.25}{0.75}{4}{5}{figures/results/comfort_point_to_select_an_interval.png} and engaging \statnew{3.92}{0.90}{4}{4}{figures/results/engagement_point_to_select_an_interval.png}.
They explained that \textit{point\&drag} was very intuitive \statnew{4.67}{0.49}{5}{5}{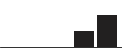} as it was \quotestudy{similar to their current practise [, i.e.,] clicking on desktops} and \quotestudy{touching on tablets}.
% \ben{Above, you need differnetiate better between preference and intuitiveness scores. Perhaps, included that into the brackets? It;s not always clear which you talk about}.
% On the other hand, \todo{some limitation: e.g. tactile feeling, occlusion, etc. point out by some participants.}
Participants also preferred \textit{cover} \statnew{3.67}{0.78}{4}{4}{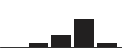},
% due to intuitiveness \statnew{4.33}{0.65}{4}{4}{figures/results/intuitive_cover_to_select_an_interval.png} and engagement \statnew{3.83}{1.11}{4}{4}{figures/results/engagement_cover_to_select_an_interval.png}.
reporting this \pInt{} to be natural and easy to perform (intuitiveness: \statnew{4.33}{0.65}{4}{4}{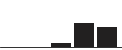}).
% For example, P1 said \quotestudy{it [cover to select an interval] is intuitive because what I hide is what I do not want to select.} 
P1 and P9 found themselves engaged and immersed in the data when performing \textit{cover}. 
P1 said \quotestudy{it [cover] is fun because I feel involved in the virtual world.}
P9 commented that \quotestudy{it [cover] is just like communicating with data using body language.}
% Interestingly, P3 strongly disagreed with selecting data by hiding it using covering because \quotestudy{I would like to actively select the area by covering, instead of covering the area that I don't want. It would greatly improve the engagement.} 

Participants rated \textit{fold} rather neutral \statnew{2.75}{0.97}{3}{2}{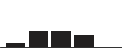} because of less comfort \statnew{2.92}{1.31}{2.5}{2}{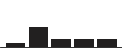}.
Reasons reported included that folding damages the paper (P4, P5, P8) and requires extra effort (P2, P3, P4, P12). P4 further commented that both hands would be required to perform the \pInt{}: \quotestudy{it is troublesome for me to use both hands to interact with the paper.} Participants 
% preferred few movements when manipulating paper sheets because
added that they were \quotestudy{sometimes lazy}.
% \ben{'troublesome sounds odd here. I am not sure what the real problem is though. Why is using both hands a in issue? Perhaps you have more information on that?}

% fold to zoom                    3.33     1.15   3.33    0.98    3.42    1.08    3.17    0.83
% point to zoom                   4.58     0.67   4.17    0.83    4.08    0.67    4.17    0.58
% translate to zoom               4.67     0.49   3.92    1.24    3.75    0.75    3.83    0.93

% fold to zoom	                3.5	    2   	3.5 	4   	3.5 	4   	3   	3
% point to zoom	                5   	5   	4   	4   	4   	4   	4	    4
% translate to zoom	            5   	5   	4   	5   	4   	4   	4   	4
\para{Zoom.} 
Participants were asked to zoom in and out the given scatterplot to get an overview or obtain detailed information about the data.
The two most preferred \pInt{}s were \textit{pinch} \statnew{4.17}{0.58}{4}{4}{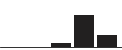} and \textit{translate} \statnew{3.83}{0.93}{4}{4}{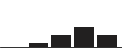}. One reason was that both \pInt{}s are intuitive (\textit{pinch}: \statnew{4.58}{0.67}{5}{5}{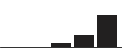}, and \textit{translate}: \statnew{4.67}{0.49}{5}{5}{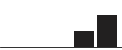}). Participants reported again that these \pInt{}s were natural and similar to daily practice. \quotestudy{I think everyone [who uses smartphones or tablets] is used to pinch,} P8 said. 
For \textit{translating} papers, P1 said \quotestudy{I do this in the real world when I want to see something larger on paper.}
% \ben{the brackets inside quotes are not usual. Is this something the participant said? I'd convert that into regular sentences.}

%\re{Although \textit{fold} scored neutral in preference \statnew{3.17}{0.83}{3}{3}{figures/results/preference_fold_to_zoom.png},
% mainly due to diverse intuitiveness \statnew{3.33}{1.15}{3.5}{2}{figures/results/intuitive_fold_to_zoom.png}. Four participants (P1, P3, P6, and P9) thought that \mapping{fold}{zoom} was not intuitive and described it as \quotestudy{unexpected} and \quotestudy{difficult to understand.}
% It was also not comfortable (similar to \mapping{fold}{select an interval}) for them to fold the paper because folding could possibly \quotestudy{damage the paper} (P3-P5), and it required more effort, \ie, \quotestudy{required to use both hands to manipulate the paper} (P1-P4, P6, and P8).
P7 found that \mapping{fold}{zoom} had a unique advantage in terms of preciseness and preferred using it because it allowed for \quotestudy{controlling the exact amount of zoom in.}
% This could also be reflected later in task 2, which two participants have changed to use \mapping{fold}{zoom}.
% As a result, we believe folding is still a potential interaction when it is matched with a suitable \op{}.

% flip to pan                     4        1.04   3.75    1.06    3.83    1.11    3.58    1.00
% point to pan                    4.75     0.62   4.08    0.90    3.92    1.00    4.33    0.78
% tilt to pan                     4.5      0.52   4.08    1.00    4.25    1.06    4.08    0.79

% flip to pan	                4   	4   	3.5 	3   	4   	5   	4   	4
% point to pan	                5   	5   	4   	5   	4   	5   	4.5 	5
% tilt to pan	                4.5 	5   	4   	5   	5   	5   	4   	4
\para{Pan.}
Participants were asked to pan a given scatterplot in different directions. Overall, \textit{point\&drag} was again strongly preferred \statnew{4.33}{0.78}{4.5}{5}{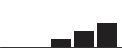} because of participants' familiarity with \pInt{}s on touchscreen (intuitiveness: \statnew{4.75}{0.62}{5}{5}{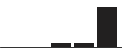}). 
It is exciting that both \textit{tilt} \statnew{4.5}{0.52}{4.5}{5}{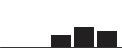} and \textit{flip} \statnew{4}{1.04}{4}{4}{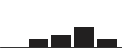} were also highly ranked. P1 emphasized that \quotestudy{flip [to pan] is surprisingly easy to understand. It's like there is a bigger visualization behind.}
% \ben{again, () are odd. if you add content yourself that the participant did not say, use []}. 
\textit{Tilt} was ranked intuitive \statnew{4.5}{0.52}{4.5}{5}{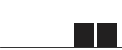} and engaging \statnew{4.25}{1.06}{5}{5}{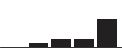}, reporting this action to be natural and playful to perform the \textit{pan} command, as well as \quotestudy{similar to a waterfall} (P2, P8), \quotestudy{playing games} (P2, P3), and \quotestudy{driving} (P1).
% However, \mapping{flip}{pan} is less comfort than \mapping{tilt}{pan} or \mapping{point\&drag}{pan} because that the interaction is \quotestudy{discontinuous} (P3 and P12) while the panning \op{} should be continuous. This led to \quotestudy{confusion} (P8 and P10). P8 said \quotestudy{flipping many times will make me confused}, which was echoed by P10 saying \quotestudy{I need to remember the previous movement.}

% collocate to link and brush     4.58     0.51   4.5     0.67    4.33    0.78    4.5     0.52
% collocate to link and brush	5   	5   	5   	5   	4.5 	5   	4.5 	5
\para{Link\&Select.} 
% Coordinate is used to investigate participants' preference to multiview \pIntFull{}s for data exploration. Participants could link different visualizations and select the same data in all linked visualizations.
Participants were asked to first select countries from a bar chart by selecting a continent in a pie chart (\autoref{fig:int_ops_illustration}(f)) and then select a time in a timeline to show different data in the pie chart about that specified time (\autoref{fig:int_ops_illustration}(c)).
\textit{Collocate\&point} was strongly preferred \statnew{4.5}{0.52}{4.5}{5}{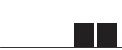} due to its naturalness and strong familiarity of selection by ``touching'' \statnew{4.58}{0.51}{5}{5}{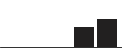}.
% and rated high in all aspects (intuitive: \statnew{4.58}{0.51}{5}{5}{figures/results/intuitive_collocate_to_link_and_brush.png}, comfort: \statnew{4.5}{0.67}{5}{5}{figures/results/comfort_collocate_to_link_and_brush.png}, and engaging: \statnew{4.33}{0.78}{4.5}{5}{figures/results/engagement_collocate_to_link_and_brush.png}).
\textit{Collate} was less preferred \statnew{3.42}{0.90}{3.5}{4}{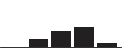} because of occlusion (P1-5, P7, P8, and P11). However, half of the participants (P1-3, P5, P8, and P12) appreciated \textit{collate} for revealing temporal changes. They described \textit{collate} as novel and engaging and that they could easily focus the changes at the top visualization.
\subsubsection*{Practical Viability and Observation}

% select an interval  zoom       pan    link and brush
% cover               fold       point  collocate         1
%                     translate  tilt   collocate         2
% point               point      point  collocate         5
%                               tilt   collocate         1
%                     translate  flip   collocate         1
%                               point  collocate         1
%                               tilt   collocate         1

% \usepackage{booktabs}

% \begin{table}
% \centering
% \caption{This table shows the initial combination of \IntOp{}s used by the participants}
% \label{tab:combination}
% \begin{tabular}{c|c|c|c|c} 
% \toprule
% \begin{tabular}[c]{@{}c@{}}\textbf{Select an}\\\textbf{Interval}\end{tabular} & \textbf{Zoom} & \textbf{Pan} & \textbf{Link\&Select} & \textbf{Count} \\ 
% \hline
% Point & Pinch & \begin{tabular}[c]{@{}c@{}}Point and\\Drag\end{tabular} & Collocate & 5 \\ 
% \hline
% Cover & Translate & Tilt & Collocate & 2 \\ 
% \hline
% Cover & Fold & \begin{tabular}[c]{@{}c@{}}Point and\\Drag\end{tabular} & Collocate & 1 \\ 
% \hline
% Point & Pinch & Tilt & Collocate & 1 \\ 
% \hline
% Point & Translate & Flip & Collocate & 1 \\ 
% \hline
% Point & Translate & \begin{tabular}[c]{@{}c@{}}Point and\\Drag\end{tabular} & Collocate & 1 \\ 
% \hline
% Point & Translate & Tilt & Collocate & 1 \\
% \bottomrule
% \end{tabular}
% \end{table}

% \ben{I don't understand the paragraph below. Can you abstract a little and say what you found? You can put the numbers in a table if you need them. }
\noindent All participants could complete the question and explore the data in Task 2.
% and\re{tried six different combinations of \IntOp{}s for \textit{zoom\&pan}. 
% In the following, numbers in brackets refer to number of participants). They are (\mapping{pinch}{zoom}, \mapping{point\&drag}{pan}) (5), (\mapping{translate}{zoom}, \mapping{tilt}{pan}) (3), (\mapping{fold}{zoom}, \mapping{point\&drag}{pan}) (2), (\mapping{pinch}{zoom}, \mapping{tilt}{pan}) (1), (\mapping{fold}{zoom}, \mapping{tilt}{pan}) (1), and (\mapping{translate}{zoom}, \mapping{flip}{pan}) (1).
% As shown Table~\ref{tab:combination}, seven combinations of \IntOp{}s were tried by 12 participants.
% Participants could perform data exploration with three different affordances: point-related \IntOp{}s (\ie, (\mapping{pinch}{zoom}, \mapping{point\&drag}{pan})), spatial \IntOp{}s (\ie, (\mapping{translate}{zoom}, \mapping{tilt}{pan}) and (\mapping{translate}{zoom}, \mapping{flip}{pan})), and mixture of point-related and spatial \IntOp{}s.
% At the end of the study, there are four participants choosing point-related only \IntOp{}s, four participants choosing spatial only \IntOp{}s and seven participants sticked with using a mixed different combinations.
During exploration, there are four participants (P3, P5, P6 and P12) changed the \IntOp{}s. 
Two participants (P5 and P12) had switched from \mapping{pinch}{zoom} to \mapping{fold}{zoom} when answering the question. While they first used \mapping{pinch}{zoom} and further \mapping{point\&drag}{pan} to the cluttered lines,
% (as shown in~\autoref{}) \todo{image}
it required several trials of zoom and pan to observe the trend, which was tedious.
Thus, they tried using \mapping{fold}{zoom} because they noticed that folding the paper might possibly zoom in to that specific area easier.
% They both could then zoom and report the trend by folding the paper once.
% Moreover, we observed point-related gestures are error-prone due to inaccurate finger detection. For example, it is easy to make selection twice and trigger zoom or pan in the reverse direction. It thus required participants more efforts to move their fingers with larger movement for the ease of detection.
Moreover, P6 and P3 found alternatives to \textit{point} actions. P6 has switched \mapping{collocate\&point}{link\&select} to \mapping{collate}{link\&select} for more accurate selection
and P3 had changed \mapping{point\&drag}{pan} to \mapping{tilt}{pan} for the free exploration in the remaining time. P3 commented that \quotestudy{it [tilt] is much easier to perform than dragging when the hand tracking is not working well.}

% \para{\re{Users' attitudes and reactions}{51}}
\subsubsection*{Users' attitudes and reactions}
\noindent All participants enjoyed interacting with the printed visualizations using \pIntFull{}s and looked forward to the complete prototype system.
% Participants are able to recall all the \IntOp{}s learned without any hints. The average recall rate (saying both the \op{} and corresponding \pInt{} correctly) is 94.7\%.
As key strengths for data exploration participants reported that paper interactions \quotestudy{increase the capability of static visualizations} (5/12 participants) and were generally \quotestudy{convenient} (4/12 participants). For example, P3 commented that \quotestudy{direct interacting with papers is more convenient than using PC for data exploration.}
Yet, the key weaknesses were reported to be \quotestudy{the durability of the paper} (4/12 participants) and \quotestudy{ergonomic issues brought by the HMD} (4/12 participants).
Overall, all participants stated that they would use these \IntOp{}s for data analysis in other contexts. Five participants would have liked to perform multiview analyses on experiment reports and academic papers.
Seven participants envisioned that paper interactions could be used in presentations to interact with data directly on the printed reports.
Moreover, four participants can see paper interactions being used in education due to the interactions engagingness.
For instance, P10 stated \quotestudy{It would be great if students could interact with the map directly to learn about geography.}
Furthermore, due to the ubiquity of paper and the ease of deploying interactions,
% \ben{what is that ease of distribution? You mean deployment of the interactions? Kind of, if people have and wear HMDs}, 
three participants imagined using it inside shopping malls and exhibitions to interact with the materials (\eg, leaflets) received.
% \ben{I don't think you need participant IDs when you refer to the quantity of people who support an argument. Just state the numbers. Identifiers are important when quoting. Again, removing these formal numbers increases readability} 
\section{Design Implications}
\label{sec:discussion}
% \ben{-------- Ben to continue here ------}

Based on the results and observations in the study, we derived \numDesign{} design implications (\ie, I1-I\numDesign{num}) for future designs and studies.
% I1-4 focus on the design of \IntOp{}s. To better facilitate \pIntFull{}, we present I5 for the visualization design and I6 for the paper design.

\para{I1. Provide Redundant Actions to \textit{Point}.}
Due to the limited accuracy of fingertip detection using only computer vision algorithms, we suggest designers provide redundant \pInt{}s, such as cover, tilt, and translate, to \textit{point} for selecting an interval, panning, and zooming, respectively.
Although point-related \pIntFull{}s are the most preferred \pIntFull{}s for all \op{}s presented in the study,
it is still challenging to provide a good and precise pointing experience similar to touchscreen due to the limitations of fingertip tracking with occluded and fast-moving fingers~\cite{bandini2020analysis} and depth estimation with deformable papers.
In addition, imprecise pointing required participants to increase their fingers' movement when they conducted data exploration using pointing gestures.
\re{}{Moreover, the study from Spindler~\etal~\cite{spindler2014pinch} showed that the task completion time of spatial input (i.e., 3D translation) for 2D document navigation on mobile phones was faster than conventional point-based input.}{link with previous work}
As a result, we suggest that designers use cover, translate, and tilt to complement pointing in range selection, zoom, and pan, respectively, with more accurate detection and similar high scores for intuitiveness, comfort, and engagement. 

\para{I2. Make Use of Different Actions for Command Shortcuts.}
Paper actions can be utilized as shortcuts to save users' efforts.
Paper sheets can provide additional \pInt{}s (\eg, flip and fold) compared to the mouse, touchscreen, and keyboard (\ie, click, touch, and keypress).
We can utilize some of the proposed \IntOp{}s (\ie, \mapping{fold}{zoom}, \mapping{flip}{pan}, and \mapping{collate}{link\&select}) that provide unique advantages as shortcuts.
For \mapping{flip}{pan}, participants agreed on its intuitiveness and showed its strength in panning a long distance, which can relieve users from pointing and dragging multiple times and tilting for a long time. 
Furthermore, \mapping{fold}{zoom} is beneficial when dealing with skewed data distribution (\ie, dense points in the corner in the visualization).
Lastly, users are engaged to use \mapping{collate}{link\&select} to quickly link two visualizations and make a selection simultaneously to focus on the temporal changes of the visualization on the top without context switching compared with \mapping{collocate\&point}{link\&select}.

\begin{figure}[t]
 \centering
 \includegraphics[width=\linewidth]{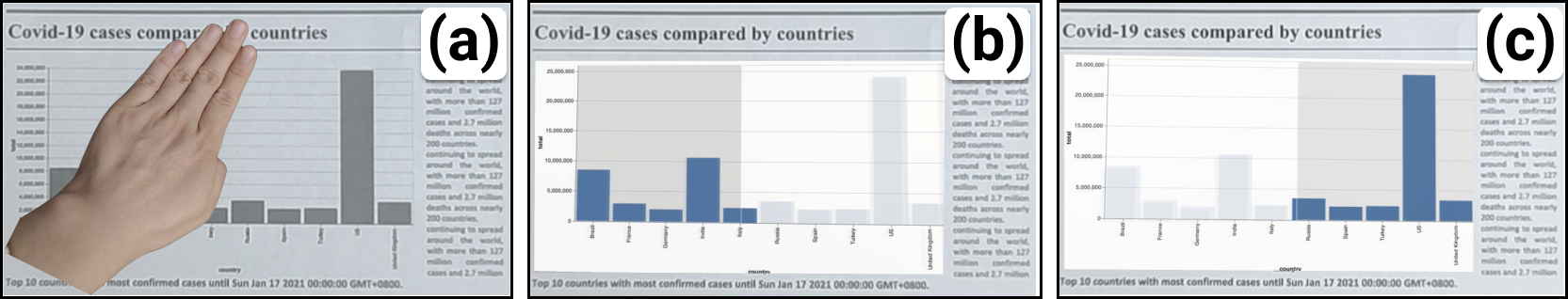}
 \caption{When a user (a) covers five bars in a bar chart, (b) the covered bars or (c) the uncovered bars can be selected.}
 \label{fig:exam_cover}
%  \vspace{-3mm}
\end{figure}

\para{I3. Support both Selection and Inverse Selection for \textit{Cover}.}
Our study suggests supporting \mapping{cover}{select-an-interval} for both selection and inverse selection.
% because these two options are exclusive. 
\textit{Cover} an area can be treated as selecting wanted data or excluding unwanted data, as shown in \autoref{fig:exam_cover}.
% In the user study, we observed that participants have their habits for these two options, as shown in \autoref{fig:exam_cover}.
% While \mapping{cover}{select-an-interval} by excluding unwanted area is intuitive, the selection is made passively, which is not engaging.
In the user study, while it is easy to use the cover gesture to hide a small set of outliers and focus on the main area of the data, it becomes difficult to cover a large portion of the visualization to select a small amount of uncovered data. This trade-off has also been stated in \cite{taher2015exploring}.
As a result, designers can provide both selection and inverse selection by using different gestures, such as palm up and palm down.
% to distinguish between selection and inverse selection for different users and usages.

\para{I4. Utilize the Semantic Meaning of Paper Actions.}
% could better link to the design space
Designers should consider the semantic meaning of the \pIntFull{} to increase the intuitiveness when designing new \IntOp{}.
Although \pIntFull{}s within the same DoI could be used for a \op{}, they provide different semantic meanings related to day-to-day usages of papers.
For example, 
% shaking acts as an universal triggers for different operations while pinning and holding has a semantic meaning of freeze and select. 
participants in the workshop preferred rubbing as a filtering action (as a rubber) or a revealing action (as a cleaner). Rotation-based \pInt{}s, \ie, tilt and rotate, correlate physics-based metaphor, such as gravity.
Moreover, moving a paper sheet back and forth has implicitly provided a zooming metaphor.
As such, \mapping{translate}{zoom} and \mapping{tilt}{pan} provide strong semantic meaning and support a strong mental model~\cite{rzeszotarski2014kinetica, jacob2008reality}, thus getting high ratings in intuitiveness.

\para{I5. Intuitiveness of Actions vs. Readability of Text.}
Designers need to consider the trade-off between the intuitiveness of the \pIntFull{} and the readability of the digital visualization.
% Reading a data visualization is different from reading images.
Users need to read both the text and the visual marks on the visualization for value retrieval and pattern recognition. 
Therefore, ensuring text readability is essential for designing \IntOp{}s.
In the study, some \IntOp{}s are intuitive and engaging.
However, they may not be optimal for readability. 
For example, \mapping{tilt}{pan} causes the text to be hard to read because the text is tilted.
Thus, designers might consider ensuring readability when using \pInt{}s that involve movements on paper, such as tilting and translating.
One possible solution is to fix some visualization components in place, such as the title, axis, and legend, while translating and tilting.
While the text is fixed at a certain distance for reading, the visual pattern could follow the movement for intuitiveness and engagement. \re{}{This implication could value beyond the domain of visualization to general physical documents.}{general to HCI}

\para{I6. Effects of Paper's Physical Properties on Actions.}
% \para{Effects of physical properties on \pIntFull{}.}
Visualizations can be printed on, \eg, books, A4-sized leaflets, and small paper cards.
During the investigation of \pIntFull{}s for data exploration, we found that the physical properties (\eg, size, weight, thickness, and physical constraints) could affect the usage of \IntOp{}s.
First, people prefer large-sized papers to perform actions with large movements or high precision.
% In the ideation workshop, we had provided two kinds of sizes for participants to interact with.
Participants in the ideation workshop preferred to interact with visualizations on a larger paper size because they can perform \pInt{}s requiring large movements easier, such as folding.
On the other hand, it becomes hard to select a large area with \mapping{cover}{select-an-interval} if the size of the paper is too large.
Second, the weight of the paper used should be light, so that none of the participants reported that interacting with the paper for about an hour was tiring.
Furthermore, participants can easily pick up the paper sheet for a better angle to view the visualization.
% However, with larger paper size, participants are hard to
Third, the thickness affects the use of \pIntFull{}.
In the user study, we used standard office paper, which is thin. While participants can easily fold the papers, it is hard to perform point-related gestures, as the thin paper cannot support the force given by the participants' fingers. Media like paper cards may provide an ideal experience for these \pInt{}s.
% Moreover, new material should be explored to increase the durability of the paper. In such, it could increase the  
% As such, to better support point-related gestures, thick papers, such as cardboard, are recommended.
Fourth, designers should consider the paper format. Paper actions (\ie, \textit{fold}, \textit{collocate}, \textit{collate}, \textit{flip}, and \textit{staple}) are constrained if papers are bounded together. For example, participants can only fold one side of the paper and cannot perform multi-\pIntFull{}s if papers are bounded as books and magazines.
\re{}{These findings might still be valid outside the field of visualization.}{general insight}
% 2) paper-\pInt{} constraint: we cannot fold the paper to increase the area of the paper.
% Lastly, physical constraints can be utilized as \pInt{} guidance. \mapping{fold}{zoom} implicitly defined the boundary of the zoom extent because we cannot unfold the unfolded paper to increase the area of the visualization to zoom out more than the original size.
\section{Discussion}
% This section discusses the usage scenarios collected from the user study of using \pIntFull{} for data exploration, limitations, and future work.

\para{Paper Interaction Design Space.}
\re{}{
Our design space captures actions and their mappings to commands. It shows the mappings we have investigated in this study while leaving the possibility and feasibility of other mappings to future work. The design space and our findings can help choosing mappings for real-life systems. It shows that for some combinations of actions and commands, multiple options exist.
% Our design space provides a structured way to help designers figure out feasible paper action and data visualization command mapping. It may miss dimensions that reflect user preference. With the user study, we found that semantics might be an important aspect that affects user preferences. It would be more engaging and intuitive if the command visually maps with the action.
}{Balance the discussion toward the design space (R1)}

\re{}{Specifically, our design space and study help making more informed decisions for creating systems based on paper interactions. For example, participants in our study preferred point-based gestures or spatial paper actions (\eg, \textit{tilting}, \textit{translating}) over \textit{folding}. Designers should also consider redundant actions for the same command, especially when a command requires different levels of granularity. For example, tilt could be used for fine-grained panning, while flip could be used to quickly pan over large distances.
% Considering integrating paper actions in a future paper-based interactive system in AR, designers could consider similar affordance for related commands. Participants most likely chose point-focused gestures or spatial paper actions (\eg, \textit{tilt} and \textit{translate}) and completed the task in task 2 of the user study. Moreover, designers could also provide more than one action for different granularity control of a data visualization command. For example, \textit{tilt} could be used for fine-grained panning, while \textit{flip} could be used for large distance panning.
}{Clarify how to integrate these interactions in an interactive system in AR (R3)}

% \para{General insights from and to HCI.}
% \re{}{Our design implications (I1 and I2) can be drawn from HCI research. Spindler~\etal~\cite{spindler2014pinch} studied that the task completion time of spatial input (\ie, 3D translation) for 2D document navigation on mobile phones was faster than conventional point-based input, which supports our implication of involving more spatial actions with paper. Moreover, our design implications could also be applied to general HCI content. Specifically, our design implication on readability (I5) and the paper properties (I6) could also help towards designing paper interaction for augmented physical documentations.}{Add a discussion point about how the user study involved mostly general HCI interactions that could as well be applied to paper, as well as to general HCI content (and not just visualization) (R2)}

\para{Possible usage scenarios.}
% In this paper, we do not show that interacting with printed visualization could replace mobile devices or desktop experience but provide a way for users to interact with static visualizations, which are available anywhere, using AR.
Our study demonstrates that interacting with printed visualizations is fun and practically viable.
% potentially can attract people's attention in data exploration in public areas, such as daily usage~\cite{mistry2009wuw}, exhibitions and presentations.
% With the rapid development of AR HMD~\cite{xiong2021augmented}, 
We envision five application opportunities for applying paper interactions with data analysis:
(1) \textbf{education}: we can add interactivity to paper sheets that could benefit classroom teaching \re{by creating an immersive learning environment, such as teaching data visualization~\cite{bae2021touching}, geography and chemistry}{that are still common to use paper and data visualization, such as teaching data visualization~\cite{bae2021touching}, geography (\eg, printed maps), chemistry (\eg, printed experiment results), and math (printed or hand drawn plots)}{Clarify the target users and usage scenarios (R3, R4)};
(2) \textbf{brainstorming sessions}:
% paper workspace is still existing, and the paperless office remains a myth~\cite{sellen2003myth}, there are still users, such as 
UX designers may consider making use of different types of papers for different tasks. In addition to qualitative data analysis with printed reports, sticky note is a common tool for supporting brainstorming~\cite{subramonyam2019affinity}. \re{Supporting interactions on various types of paper in use reduce the context switching between the real and augmented environment}{Supporting interactions directly on papers can reduce the context switching between the desktop visual analytics tools and sticky notes~\cite{subramonyam2022composites}}{Clarify the target users and usage scenarios (R3, R4)};
(3) \textbf{exhibitions/presentations}: audiences could directly interact with paper handouts (\eg, worksheets, leaflets, and pamphlets; as shown in \autoref{fig:teaser}) provided \re{}{without switching back and forth between mobile phones and handouts}{Clarify the target users and usage scenarios (R3, R4)} to seek more information during visiting an exhibition and attending a presentation.
% (4) \textbf{public area showcase}: since this interface is new and intuitive, it might catch people's attention on data when displaying data in a public area.
% Instead of picking out the phones from pockets to retrieve more information online, this type of interface allows users to quickly investigate the information directly from the leaflet or paper provided.
% For example, a newspaper is distributed on the streets to make people aware of the seriousness of Covid-19. People could point to the newspaper to link and select specific data points (\autoref{fig:exam_multiview}), similar to \cite{mistry2009wuw}. Or they could quickly fold the newspaper to select the information that is important to them.
(4) \textbf{collaboration}: \re{with intuitive \pIntFull{}s, people could quickly explore the data printed on the paper sheets in different scenarios, such as university advertising on the school open day (\autoref{fig:teaser}), promotion of products, and exhibition, for data exploration and decision making. Moreover, supporting short analytical sprints could help collaboration between diverse domains~\cite{ens2020uplift}.}{with intuitive \pIntFull{}s, people could quickly explore the data printed on the paper sheets, which can support short analytical sprints. It helps enhancing collaboration between diverse domains~\cite{ens2020uplift}.
(5) \textbf{AR-based authoring tools for interactive visualization}: current AR-based authoring tools~\cite{chen2020augmenting,DBLP:journals/tvcg/ChenS0WQW20,cordeil2019iatk,sicat2018dxr} only support minimal or even no interaction configuration, which hinders users to interact with the created visualization. Our paper interaction design space could help developers and researchers to further extend their tools to support feasible paper interactions for interactive AR visualizations.}{Clarify the target users and usage scenarios (R3, R4).}

% \begin{figure}[t]
%  \centering
%  \includegraphics[width= 1.0\linewidth]{figures/example_multiView.png}
%  \caption{An example of multiview analysis: (a) A newspaper reports Covid-19 confirmed cases with a pie chart, a line chart, and a choropleth map. (b) Digital layers of the three charts are augmented to the newspaper with our prototype. (c) A tooltip appears when pointing to a country, and meanwhile, its corresponding line and the sector of the continent are highlighted in the line chart and pie chart.}
%  \label{fig:exam_multiview}
% \end{figure}

\para{Study Limitations.}
Despite our best effort, this study has some limitations to be aware of.
First, our investigation was mainly based on the workshop's outcomes and unable to investigate all possible designs conclusively and exhaustively.
% More studies could be done to expand the design space for analysis process \& provenance~\cite{heer2012interactive}.
Second, AR technology is still premature.
For example, the inaccurate detection of fingers and paper hindered the user experience; 
blurry text, due to the fixed focal length of HoloLens 2, caused eye fatigue and strain (reported by half of the participants).
Furthermore, as an exploratory study, we have implemented a subset of 11 \IntOp{}s \re{}{with two paper-specific actions (\ie, \textit{cover} and \textit{fold})}{Some key actions are not implemented and studied in the user study (R2)} to complete simple tasks. \re{}{The sample of the user study is also small}{the evaluation is not strong enough and can be improved} and did not consider the analytic benefits of \IntOp{}s.
However, our design space with \numComb{} \IntOp{}s, our grouping of \pIntFull{}s and \op{}s, as well as our study results provide a good framework for a more systematic exploration of paper \IntOp{}s in the future.

\para{Future Work.}
% In this work, we have mainly focused on the \IntOp{}s for view manipulation. There are more complicated \op{}s, such as data \& view specification (\ie, derive, filter, sort, and encode) and analysis process and provenance (\ie, record, annotate, share, and guide), which are important for users who want to create and share their data insights using visualizations~\cite{heer2012interactive}.
More studies could be done to expand the design space for the analysis process \& provenance~\cite{heer2012interactive}.
The prototype could also be extended to conduct more studies for assessing other aspects (\eg, task accuracy and completion time for analytic benefits and memory test for intuition), as well as complex tasks for authoring visualizations and immersive collaborative analysis~\cite{billinghurst2018collaborative,ens2020uplift, DBLP:journals/tvcg/ChenYCXZQW22}. \re{}{It could further include more paper-specific actions to support more commands, such as \mapping{dogearing}{pin-view}.}{Some key actions are not implemented and studied in the user study (R2)}
Furthermore, artificial intelligence could be introduced to facilitate better interaction support in AR~\cite{wu2021ai4vis, wu2022computableviz}, and better paper detection and finger detection with depth cameras and extra sensors.
% future work could investigate advanced algorithms utilizing the depth camera for paper detection and extra sensors for finger detection to improve the usability of the prototype.
Last but not least, it is interesting to explore the possibility of using \pIntFull{}
as metaphors for intuitive gesture design in the air (without actually interacting with physical papers) to interact with data visualizations in virtual reality (VR)~\cite{DBLP:journals/tvcg/ChuXYLXYCZW22, DBLP:journals/tvcg/YeCCWFSZW21}.
Although the haptic feedback might be lost, there are more design choices when deploying paper \IntOp{}s without the physical paper. For instance, undoing a tearing action on a virtual paper sheet is possible in VR.
% \re{}{Future studies could be done to compare different VR-based methods with this VR paper-based interaction design to confirm the analytic benefits of paper-based interactions.}{neither justify nor evaluate the alternative designs like VR-based methods (R4)}
\section{Conclusion}
% \label{sec:conclusion}

This paper explores the use of paper sheets as a new means for interaction with data visualization.
We first conducted an ideation workshop with 20 VIS and HCI researchers to solicit \numComb{} \IntOp{}s. Furthermore, we construct a three-dimensional design space (\ie, \dimOne{}, \dimTwo{} and \dimThree{}) to describe and create possible \IntOp{}s and verify the feasibility of \IntOp{}s.
Lastly, we built a proof-of-concept prototype and conducted a user study with 12 participants to provide initial insights by evaluating 11 \IntOp{}s.
Our findings show that all participants considered these \IntOp{}s intuitive and engaging.
Based on the findings, we developed \numDesign{} design implications.
We found strong affordances for some interactions, physical limitations and properties of paper as a medium, cases requiring redundancy and shortcuts, and other implications for design.
% Our work demonstrates the benefit of interacting with data visualizations on paper, and our design space helps researchers and designers to design \IntOp{}s for data exploration.
We hope that our work can inspire future work on developing \IntOp{}s for data exploration more intuitively and engagingly.

% \newpage

\acknowledgments{
This work is partially supported by Hong Kong RGC GRF Grant (No. 16210321).}

\newpage
\bibliographystyle{abbrv-doi}
\balance{}
\bibliography{template}
\end{document}